\documentclass[prd,aps,floats,12pt,nofootinbib]{revtex4}
\usepackage{times,graphicx,amssymb, amsmath, natbib}
\usepackage[T1]{fontenc}

\usepackage{epsfig}

\newcommand{\beq}{\begin{equation}}
\newcommand{\eeq}{\end{equation}}

\newcommand{\ber}{\begin{eqnarray}}
\newcommand{\eer}{\end{eqnarray}}

\def\doi{http://doi.org}

\def\beq{\begin{equation}}
\def\eeq{\end{equation}}

\def\ber{\begin{eqnarray}}
\def\eer{\end{eqnarray}}

\begin{document}

\title{Interacting Quintessence and growth of structure}

\author{Azam Hussain}
\email{azamhussain@ctp-jamia.res.in} 
\affiliation{Centre for Theoretical Physics, Jamia Millia Islamia, New Delhi-110025, India}

\begin{abstract}
In standard cosmologies, dark energy interacts only gravitationally with dark matter. An extension to this picture is interacting quintessence (IQ) model  where scalar field coupled directly to cold dark matter. The percentage deviation is studied in IQ model wrt  $\Lambda$CDM  for varied values of interacting parameter W. We investigated the effect of interaction on matter , kaiser and galaxy power spectrums. Deviation in power spectrum increases with interaction on both large and small scales. On small scale, variation is comparatively smaller than on large scale. On large scale ,it is due to dark energy perturbation while it is background evolution that causes a difference on small scale. These variations decreases with increase in redshift. Herein thawing class of model with linear potential is studied. 
\end{abstract}

\maketitle
\date{\today}

\section{Introduction}
The late time acceleration in the standard Einstein gravity is propelled  by a mysterious energy component which consist of a  huge negative pressure that expands the Universe. This is called  dark energy  \cite{SN,SN2,SN3,per}. Keeping in mind the standard cosmological model, the DE undertake the simplest form of cosmological constant $\Lambda$, which has absolutely no spatial fluctuations but a negative pressure and constant energy density which covers the entire expansion history of the Universe. That leads to $\Lambda$CDM model.  Cosmic Microwave Background (CMB) \cite{ade} , Supernova Type-Ia (SnIa)  (SnIa) \cite{jla}, Baryon Acoustic oscillation (BAO) measurement in galaxy surveys \cite{bao} : can be demonstrated via $\Lambda$CDM model . However, it gets into serious conceptual problems like " fine tuning problem " \cite{win} and "coincidence problem" \cite{hu,de}. Recent observational results also indicate inconsistency with $\Lambda$CDM model \cite{discrep,riess,kids,valen}.

In order to solve these problems, a number of scalar field models
including quintessence \cite{Ratra:1987rm,Peebles:1987ek,Copeland:2006wr,Sahni:1999gb,
	Frieman:2008sn,Padmanabhan:2002ji,Padmanabhan:2006ag,
	Sahni:2006pa,Peebles:2002gy,Perivolaropoulos:2006ce,
	Sami:2009dk,Sami:2009jx,Sami:2013ssa}, phantom
field \cite{Caldwell:1999ew,Caldwell:2003vq,Carroll:2003st,Singh:2003vx,Hao},
rolling tachyon \cite{Sen:2002nu,Sen:2002in,Sen:2003mv,Mazumdar:2001mm} and others have been proposed. Another way to look at this problem is through interaction between dark energy and other matter species in the Universe
as suggested by \cite{Wetterich,Amendola,Farrar,Gubser,Farrar2}. In relation to this, minimally coupled dynamic scalar field can be extended to interacting Quientessence  models (from now IQ). In these categories of models, DE is coupled to matter (dark matter and baryons), but  coupling of DE with baryonic particles will result in  time variation of constants of nature and hence, is tightly bound by observations \cite{Hagiwara}. However, these constraints does not apply to interaction between dark sector. This interaction between DM and DE is still permitted by observations \cite{Damour,Casas}. In this work, we will focus on coupling between dark sector only.

Various types of interaction between DE and CDM have been proposed 
and investigated in the literature, like, Linear coupling of scalar field with matter (DM and baryon) suggested by Amendola \cite{Amendola}, Coupling with dark matter only \cite{Damour,Casas}, Nonlinear couplings \cite{Nonlinear1,Nonlinear2}, Variational approach \cite{Boehmer1,Boehmer2}.
In relation to aforementioned models, the coupling describes the exchange of energy-momentum between dark energy and dark matter.

Interaction between dark sector could change the expansion history of the Universe . Interaction can change source term of Poisson equation  through scalar field perturbation. 
Additionally it can produce a fifth force between matter particles which results into stronger clustering of matter \cite{NL4,NL5}. All these factors could affect structure formation on both small and large scales.
In recent years, the impact of interaction on  linear growth of structure \cite{L1,L2,L3,L4,L5,L6,L7} , non-linear structure formation \cite{NL1,NL2,NL3,NL4,NL5,NL6} has been studied in detail. Observational data like CMB , BAO, LSS, Weak lensing \cite{C1,C2,C3,C4,C5}, local gravity tests \cite{LG1,LG2}  and high redshifted intergalactic medium \cite{IGM1} have constrained the interaction, but none of them have  ruled out the interaction.

  In order to advance the work by Bikash et al.\cite{Bikash} we shall  include coupling between Quintessence and CDM and investigate its effect on linear structure formation.We consider a thawing class of model. We include red  shift space distortion term  and  GR effects in power spectrum. We vary the interaction strength to show the  percentage deviation wrt  $\Lambda$CDM model at different redshifts. We normalize IQ model to give it parameters ($\Omega_{m0}$ , $H_{0}$) that are similar to that of ordinary non-interacting Quintessence at present redshift.

The outline of the paper is : Sec II deals with  background IDE models while Sec section III discusses perturbed DE models . Linear structure formation is discussed in section IV. Result and conclusion are discussed in section V.
\\
\section{Background evolution with scalar field coupled with DM}
We have considered a model which allow coupling between dark sector. Scalar field is coupled to dark matter but not to baryons. Here, we follow prescription discussed many times in literature \citep{Amendola,Amendola2004,Amendola2008,L2,NL2,Piloyan,Sumit}. We also follow the same formalism here. The important equations are as follows:
\begin{eqnarray}
\ddot{\phi}+\frac{dV}{d\phi}+3H\dot{\phi}=C(\phi)\rho_{d} \nonumber \\ 
\dot{\rho}_{d}+3H(\rho_{d})=-C(\phi)\rho_{d}\dot{\phi} \\
\dot{\rho_b}+3H(\rho_b)=0 \nonumber\\
H^2=\frac{\kappa^2}{3}(\rho_b+\rho_{d}+\rho_\phi) \nonumber \\
\end{eqnarray}
\begin{equation}
1 = \frac{\kappa^2\rho_{b}}{3H^2}+\frac{\kappa^2\rho_{d}}{3H^2}+\frac{\kappa^2\dot{\phi}^2}{6H^2}+
\frac{\kappa^2V(\phi)}{3H^2}
\end{equation}
Herein, $C(\phi)$ comprises an interaction between dark sector. Due to lack of detail of the nature of interaction we consider it to be constant \citep{Amendola,Amendola2004} . We can even study uncoupled case by putting  $C=0$.

Now, we introduce these  dimensionless parameters:
\begin{align}
x	= \frac{\kappa\dot{\phi}}{\sqrt{6}H}, \hspace{1mm}
y	= \frac{\kappa\sqrt{V(\phi)}}{\sqrt{3}H} \nonumber \\
s	=\frac{\kappa\sqrt{\rho_b}}{\sqrt{3}H}, \hspace{1mm}
\lambda	=	\frac{-1}{\kappa{V}}\frac{dV}{d\phi} \hspace{1mm}
\Gamma	=	\frac{V\frac{d^2V}{d\phi^2}}{\left(\frac{dV}{d\phi}\right)^2} 
\end{align}

\noindent
Here, the variable  $\Gamma$ shows  potential.
In terms of $x$ and $y$, $\Omega_{\phi}$ and equation of state $w_{\phi}$ are :

\begin{align}
\Omega_\phi	=&	x^2+y^2 \\
\gamma	=&	1+w_\phi	=	\frac{2x^2}{x^2+y^2}
\end{align}

\noindent
Using aforementioned dimensionless variables, eq (1) and (2) can be converted in following autonomous systems:
\begin{align}
\Omega_\phi'	=&	W\sqrt{3\gamma\Omega_\phi}(1-\Omega_\phi-s^2)+3\Omega_\phi(1-\Omega_\phi)(1-\gamma) \nonumber \\
\gamma'	=&	W\sqrt{\frac{3\gamma}{\Omega_\phi}}(1-\Omega_\phi-s^2)(2-\gamma)+\lambda\sqrt{3\gamma\Omega_\phi}(2-\gamma)\nonumber\\
- &3\gamma(2-\gamma) \nonumber \\
s'	=&	-\frac{3}{2}s\Omega_\phi(1-\gamma) \nonumber \\
\lambda'	=&	\sqrt{3\gamma\Omega_\phi}\lambda^2(1-\Gamma),
\end{align}
here, $W=\frac{C}{\kappa}$. 
\subsection{Initial conditions to solve Background equations}
To solve this autonomous system (7) we would need initial conditions for ($\gamma$, $\Omega_{\phi}$,s,$\lambda$). We settled our initial conditions at ($z = 1000$). At ($z = 1000$) for thawing class of  models, scalar field is frozen thus  $\gamma_{i} \approx 0$.  We have taken initial value of  $\lambda_{i}$ as a model parameter.  We settles  the initial condition for
$\Omega_{\phi }$ by fine tuning it so as to obtain appropriate value of $\Omega_{\phi }$ today at $z=0$ . In the same way, we set initial value of $s$  to get appropriate value of 
 the $\Omega_{b}$ today. Subsequently, we set initial condition for IQ model to get same  $\Omega_{m0}$ and $H_{0}$ as in non-interacting case. 
\subsection{Behaviour of Background cosmological parameter}
Set of equations (7) will be solved using the earlier mentioned initial conditions and the cosmological parameters for different values of interacting parameter W will be studied. We focus on linear potential.

 Figure 1 displays the variation of equation of state as function of redshift for different values of W. Since we are considering thawing model, $w_{\phi}$ starts from -1 at $z= 1000$ for non interacting case. On adding interaction, $w_{\phi}$ increases which shows strong dark energy effect.
 
 Figure 2 displays the percentage change in matter density parameter wrt LCDM. Here percentage change is negative which implies suppression wrt LCDM. The background dark matter density ($ \Omega_{m} $) decreases wrt LCDM.
 
  Figure 3 displays the percentage change in normalized hubble parameter  wrt LCDM. Here, percentage change is positive which implies enhancement wrt LCDM. On increasing Interacting parameter (W), percentage change increases which entails  stronger dark energy effect.
 \begin{center}
 	\begin{figure*}
 		\begin{tabular}{c@{\quad}c}
 			\epsfig{file=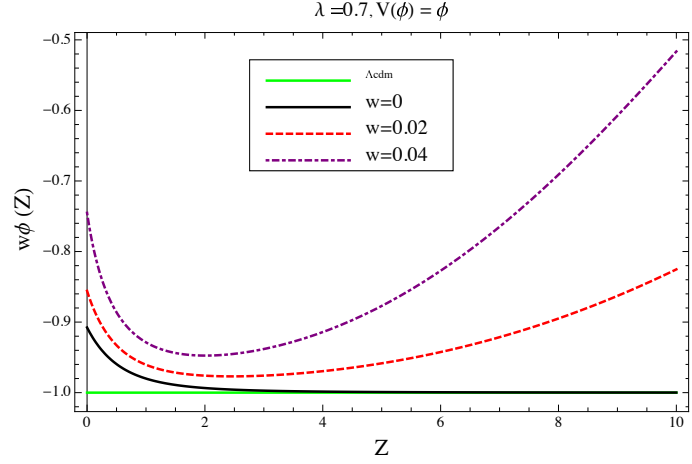,width=9.0 cm}
 		\end{tabular}
 		\caption{Conduct of the eos for different values of interacting parameter W.
 		}
 	\end{figure*}
 \end{center}
\begin{center}
	\begin{figure*}
		\begin{tabular}{c@{\quad}c}
			\epsfig{file=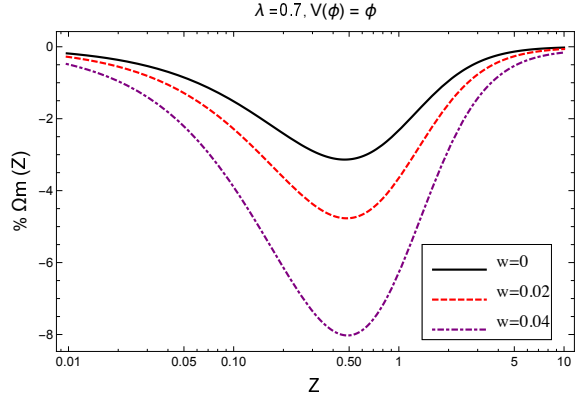,width=7.5 cm}
	
		\end{tabular}
		\caption{Percentage diversion of $ \Omega_{m} $ in IQ as compared to $ \Lambda$CDM model for different values of interacting parameter W: negative values
			in y-axis means they are less than that that of $ \Lambda $CDM. $\Omega_{m0} = 0.28$ and $\lambda_{i} = 0.7$ in these plots. Subsequently , $\% \Delta X = (X^{de} / X^{\Lambda} - 1)\times 100$.
		}
	\end{figure*}
\end{center}

\begin{center}
	\begin{figure*}
		\begin{tabular}{c@{\quad}c}
			\epsfig{file=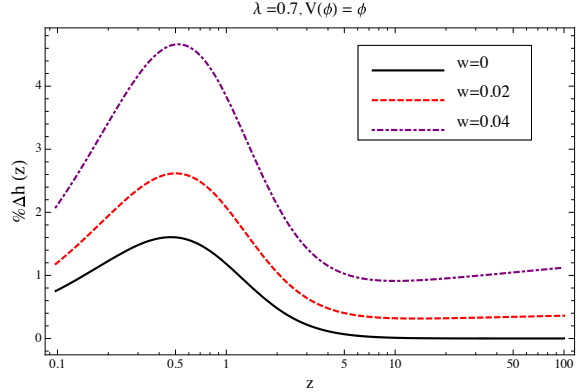,width=7.5 cm}
			
		\end{tabular}
		\caption{Percentage deviation in normalized hubble h from $ \Lambda$CDM model for different values of interacting parameter W..
		}
	\end{figure*}
\end{center}

\section{GROWTH OF LINEAR PERTURBATION WITH INTERACTING  Quintessence}
Herein effect of the interaction on matter and scalar field perturbation in linear regime will be reflected upon. Matter include both dark matter and baryons. Since dark matter perturbation is dominant and baryon follows dark matter perturbation, hence  baryonic perturbations in our study can be excluded . Even inclusion of that will not effect our result.
\\
The perturbed  (FRW) metric in conformal and Newtonian gauge is given by\\
\begin{equation}
ds^{2}=a^{2}[-(1+2\Phi)d\tau^{2}+(1-2\Psi)dx^{i}dx_{j}]
\end{equation}
where, $\Phi$ and $\Psi$ are gravitational potential and $\tau$ is conformal time. Without anisotropic stress the two can be related as $ \Phi = \Psi $.\\
Consider two components, a scalar field  and cold dark matter  described by the energy- momentum tensors $T_{\mu\nu(\phi)}$ and $T_{\mu\nu(d)}$.
Conservation equations with interacting terms for scalar field and cold dark matter as discussed by Amendola \cite{Amendola} are.
\begin{equation}
\nabla_{\mu}T^{\mu}_{(\phi) \nu}= C T_{d} \nabla_{\nu}\phi
\end{equation}
\begin{equation}
\nabla_{\mu}T^{\mu}_{(c) \nu}= -C T_{d} \nabla_{\nu}\phi
\end{equation}
here, C defines interaction between scalar field and dark matter. We assume phenomenologically, C to be constant.\\
For perturbation equations  we follow the set up as provided by Bikash et al .We generalize the results in \cite{Bikash} for the interacting scenario. We mention only relevent equations here , for detailed calculation see eg \cite{Bikash}.
\\
Relativistic poisson equation is.
\begin{equation}
\bigtriangledown^{2}\Phi-3 {\cal {H}}(\Phi^{'}+{\cal {H}}\Phi)= 4\pi G a^{2} \Sigma\delta\rho_{i}
\end{equation}
The scalar field influences a change upon source term of Poisson equation . \cite{NL4,NL5}.Here prime is derivative wrt conformal time.

The perturbation equations for scalar field coupled to dark matter is
\begin{equation}
\delta\phi^{..}+3H\delta\phi^{.} - 4\Phi^{.}\phi^{.} + 2\Phi V_{\phi} +V_{\phi \phi} \delta\phi - \frac{1}{a^{2}} \nabla^{2} \delta\phi = C(\delta\rho_{\phi} + 2\Phi\rho_{\phi})
\end{equation}
\\
where,  $V_{\phi}=\dfrac{dV}{d\phi}$, $V_{\phi \phi}=\dfrac{d^{2}V}{d^{2}\phi}$  and $\delta\phi$ is scalar field fluctuation.

Then we construct following dimensionless variables.
\begin{eqnarray}
&& g = -\frac{V_{\phi}}{H \phi \dot{•}}\\
&& q = \frac{\delta\phi H}{\phi \dot{•}}\\
&& B = 3 + \frac{H\dot{•}}{H^{2}}\\
&& B_{\phi} = 6g - \frac{1}{H}\frac{dB}{dt} - 2Bg + \frac{k^{2}}{a^{2}H^{2}}\\
&& W = \frac{C}{k}\\
\end{eqnarray}
In terms of these dimensionless parameters eq (11)  and (12) can be written as 
\begin{eqnarray}
 && \frac{d^{2}\Phi}{dN^{2}}+(1+B)\frac{d\Phi}{dN} + (2B-3+3x^{2})\Phi = 3x^{2}[\frac{dq}{dN} + \frac{3W\Omega_{d}q}{\sqrt{6}x} + (2g-B)q]\\
&&\frac{d^{2}q}{dN^{2}}+ (\frac{\sqrt{6}W\Omega_{d}}x + 2g - B)\frac{dq}{dN} + (\frac{9W\Omega_{d}}{\sqrt{6}x} - 3\Omega_{d}W^{2} + B_{q})q = 4\frac{d\Phi}{dN} + 2\Phi g + \frac{3W\Omega_{d}}{\sqrt{6}x}(\delta_{d}+2\Phi)\nonumber\\
\end{eqnarray}
Matter density contrast is given by
\begin{equation}
\delta_{d} = -\frac{2}{\Omega_{d}}[\frac{d\Phi}{dN} + (1-x^{2}+\frac{k^{2}}{3 {\cal {H}} ^{2}})\Phi + x^{2}(\frac{dq}{dN}-B_{q}+\frac{3W\Omega_{d}q}{\sqrt{6}x})]\\
\end{equation}
where, $N = lna$ \\
Here, comoving density contrast 
\begin{equation} 
\Delta_{d} = \delta_{d} + y_{d}
\end{equation}
where,
\begin{equation}
y_{d} = 3 {\cal {H}}v_{d} = \frac{2}{\Omega_{d}}(\frac{d\Phi}{dN} + \Phi -3x^{2}q)
\end{equation}
On subhorizon scale  $ \Delta_{d} \simeq \delta_{d} $ , that is usually used in study of small-scale structure (Newtonian limit). On large scales, $\Delta_{d}$ should be used instead of  $\delta_{d}$ \cite{Duniya2013}.

For growth function, we use equations given by Duniya et al  in \cite{L7} and probe the large - scale structure of the universe at different redshifts.
In the following equations  $\Phi_{d}$ represents  gravitational potential at decoupling.\\
Gravitational potential growth function $D_{\Phi}$ is defined by
\begin{equation}
D_{\phi}(k,z)=\frac{1}{(1+z)}\dfrac{\Phi(k,z)}{\Phi_{d}(k)}
\end{equation}
 $D_{d}$ is the growth function of the comoving matter overdensity.\\
\begin{equation}
D_{d}(k,z)=-\dfrac{3\Delta_{d}(k,a)\Omega_{d0}H_{0}^{2}}{2k^{2}\Phi_{d}(k)}
\end{equation}
Matter velocity growth function
\begin{equation}
Dv_{d}(k,z)=-\dfrac{3v_{d}(k,z)\Omega_{d0}H_{0}^{2}}{2\Phi_{d}(k)}
\end{equation}
Dark energy velocity growth function $D_{\phi}(k,z)$ is associated to that of dark matter by
\begin{equation}
\frac{Dv_{\phi}(k,z)}{Dv_{d}(k,z)}= -\frac{\Omega_{d}}{(1-\Omega_{d})(1+w)}(\frac{HD_{\Phi}^{'}}{Dv_{d}}+1)
\end{equation}
Ratio of comoving matter density $\Delta_{d}$ and gravitational potential $\Phi(0,k)$ can be defined as
\begin{equation}
D_{d\Phi}(z,k)=\dfrac{\Delta_{d}(z,k)}{\Phi(0,k)}
\end{equation}
here, prime is derivative with respect to redshift z.\\
Quantity f which is related to velocity perturbation and hence redshift-space distortion \cite{Duniya2016} term can be defined as 
\begin{equation}
f=\frac{Dv_{d}}{{\cal {H}}D_{d}}
\end{equation}
which reduces to the growth rate of dark matter in standard uncoupled DE m0dels.\\
Standard matter power spectrum
\begin{equation}
P(z,k) = A k^{n_s-4} T^2(k)(1+z)^{2}(\frac{\Delta_{d}(z,k)}{\Phi(0,k)})^{2}
\end{equation}
Here $A$ represents the normalization constant the value of this is determined  by $\sigma_{8}$ normalization whereas  $n_{s}$ is defined as spectral index , $T(k)$  as given by Einstein and Hu is the transfer function \cite{Eisenstein1998}.

\subsection{Initial conditions}     

To solve perturbation equations (19), (20), we need initial conditions for   ($\Phi, \dfrac{d\Phi}{dN}, q, \dfrac{d q}{dN} $) .At the time when matter dominated Universe there was negligible dark energy contribution. We have settled set our initial condition at decoupling ($z=1000$). Due to negligible dark matter initial conditions for interacting and non interacting model is same. Thus we have used initial conditions discussed by Bikash et al \cite{Bikash} for non-interacting model..

There is no contribution from DE at $z=1000$ hence scalar field perturbation is insignificant and  $q= \dfrac{d q}{dN} = 0$ .

Since  $\Phi$ is constant during matter domination,  hence $\dfrac{d \Phi}{dN}=0$ . Using Poisson equation (9) and the fact that  $\Delta_{m} \sim a$ , its initial condition is
\begin{equation}
 \Phi_{i} = \frac{-3}{2}\frac{{\cal {H}}^{2}_{in}}{k^{2}}a_{in}
\end{equation}

\begin{center}
	\begin{figure*}
		\begin{tabular}{c@{\quad}c}
			\epsfig{file=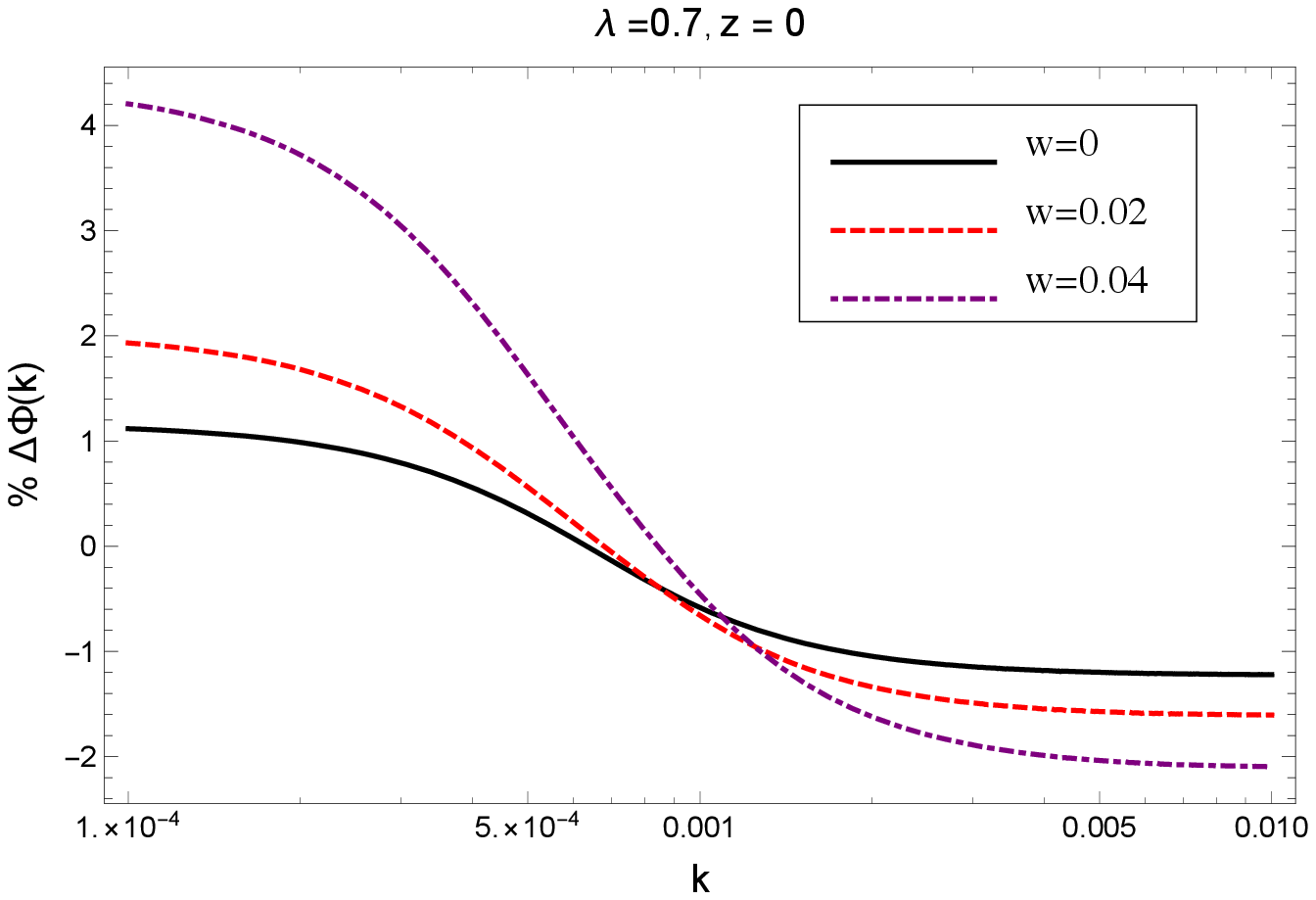,width=7.5 cm}
			\epsfig{file=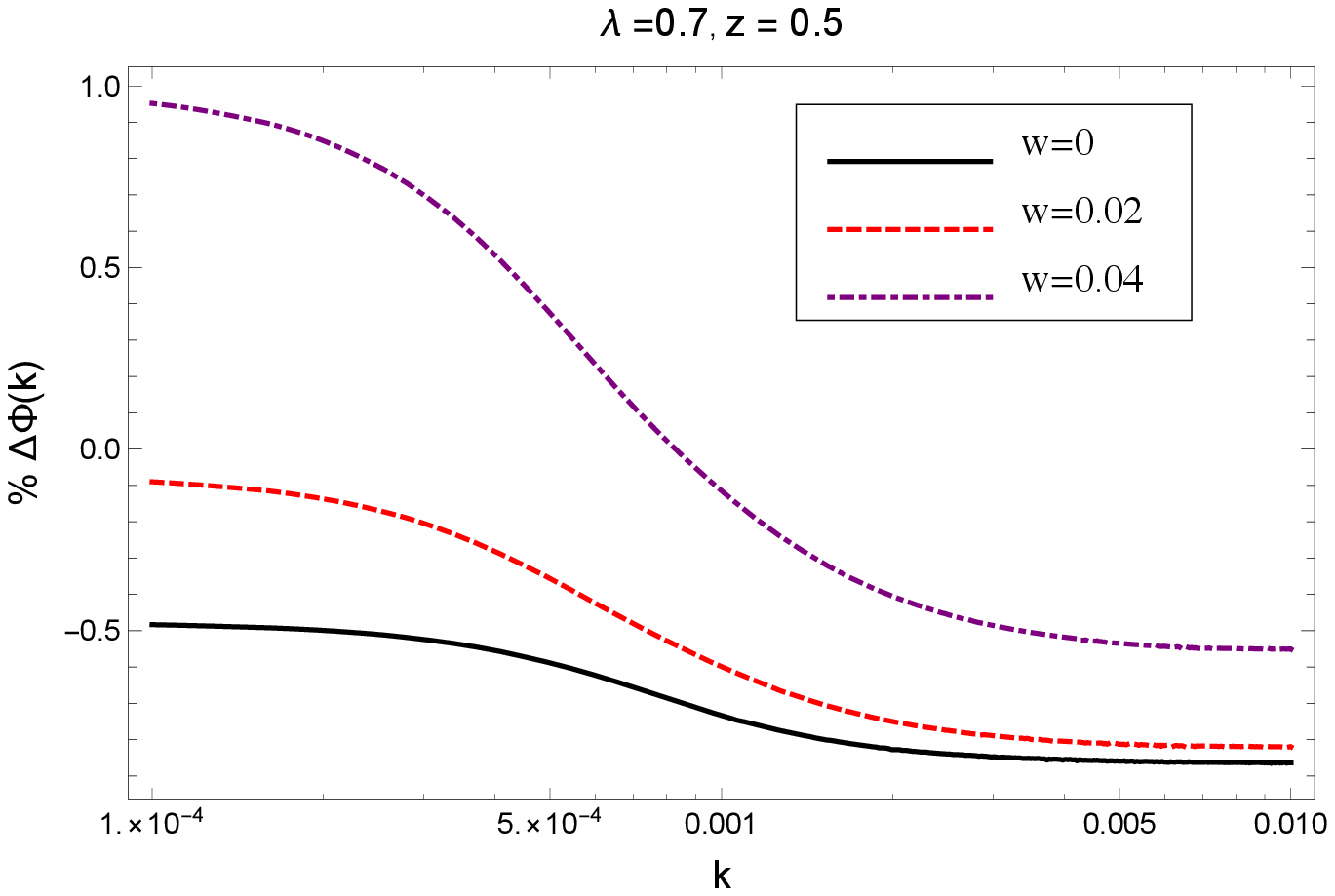,width=7.5 cm}\\
			\epsfig{file=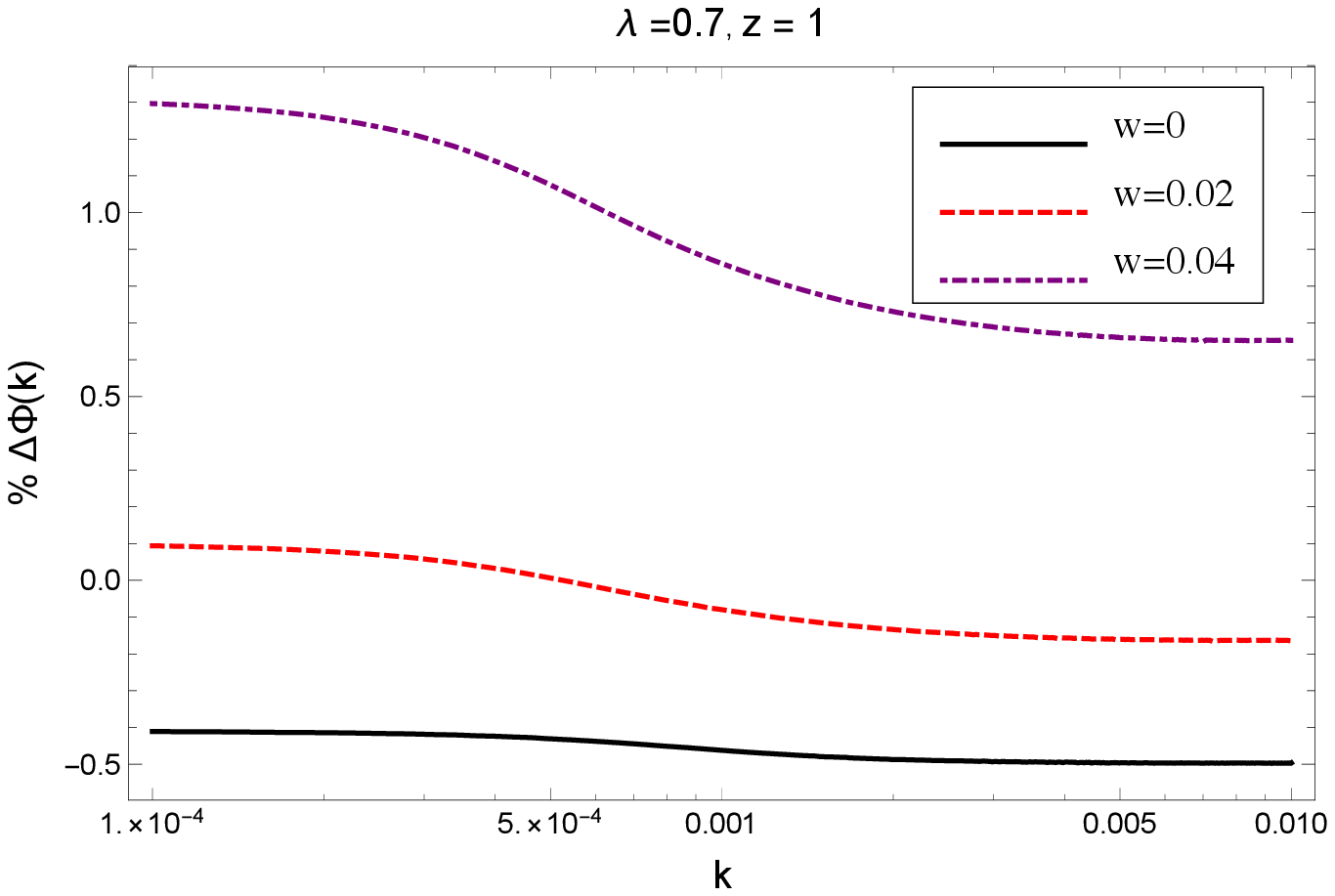,width=7.5 cm}
			\epsfig{file=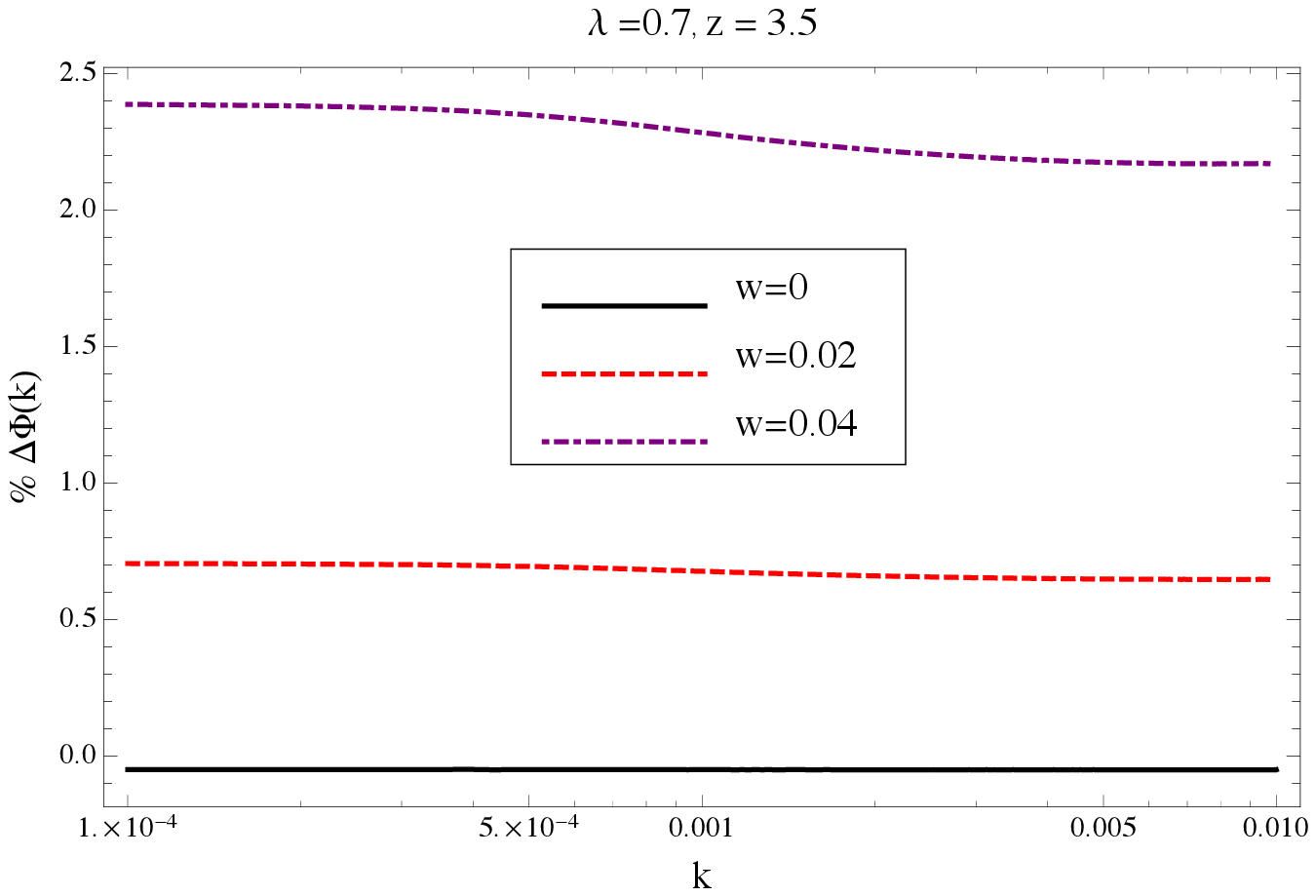,width=7.5 cm}
		\end{tabular}
		\caption{Percentage diversion  of $ \Phi $ in IQ model as compared to  $ \Lambda$CDM model for different values of interacting parameter W. 
		}
	\end{figure*}
\end{center}

\begin{center}
	\begin{figure*}
		\begin{tabular}{c@{\quad}c}
			\epsfig{file=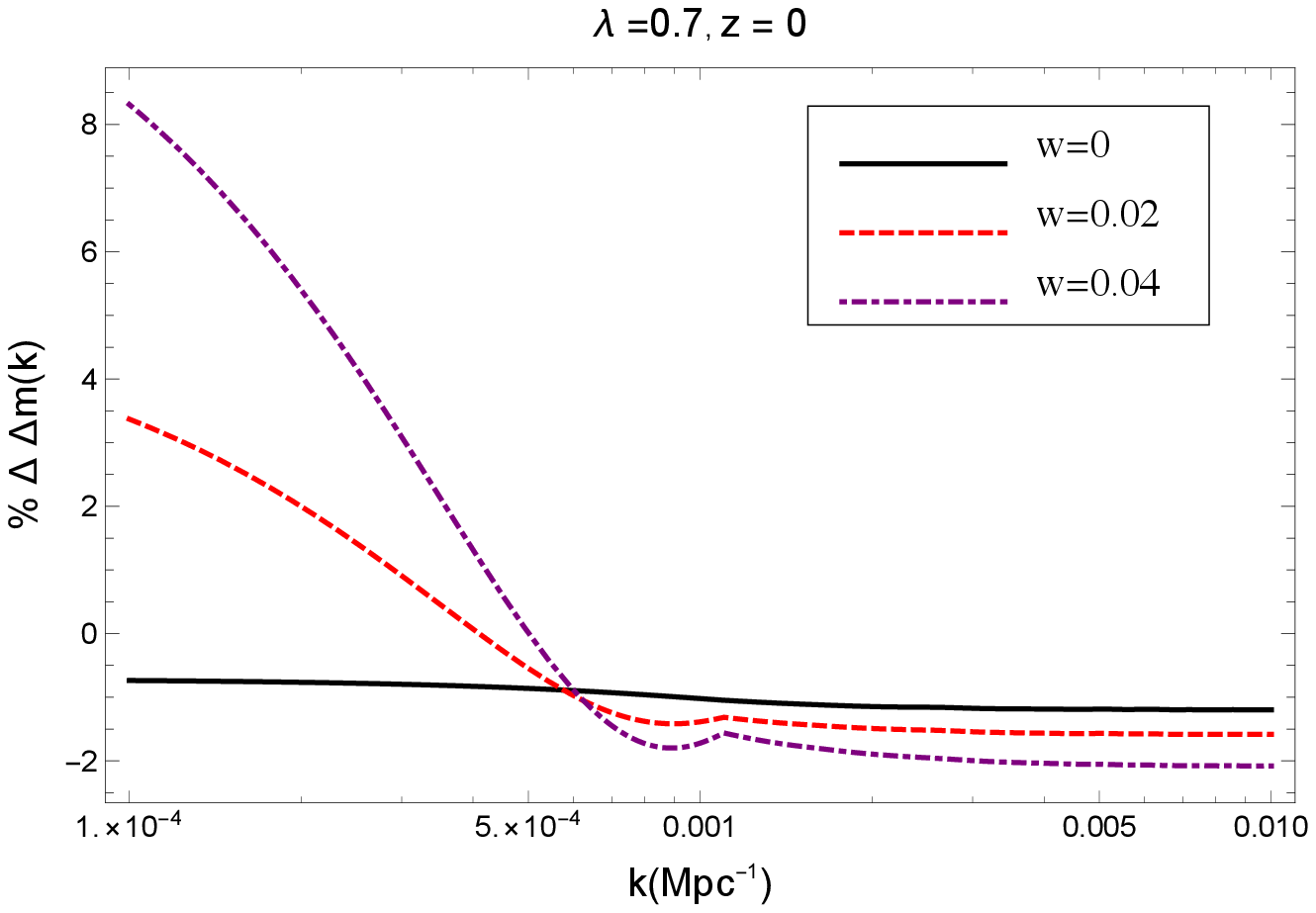,width=7.5 cm}
			\epsfig{file=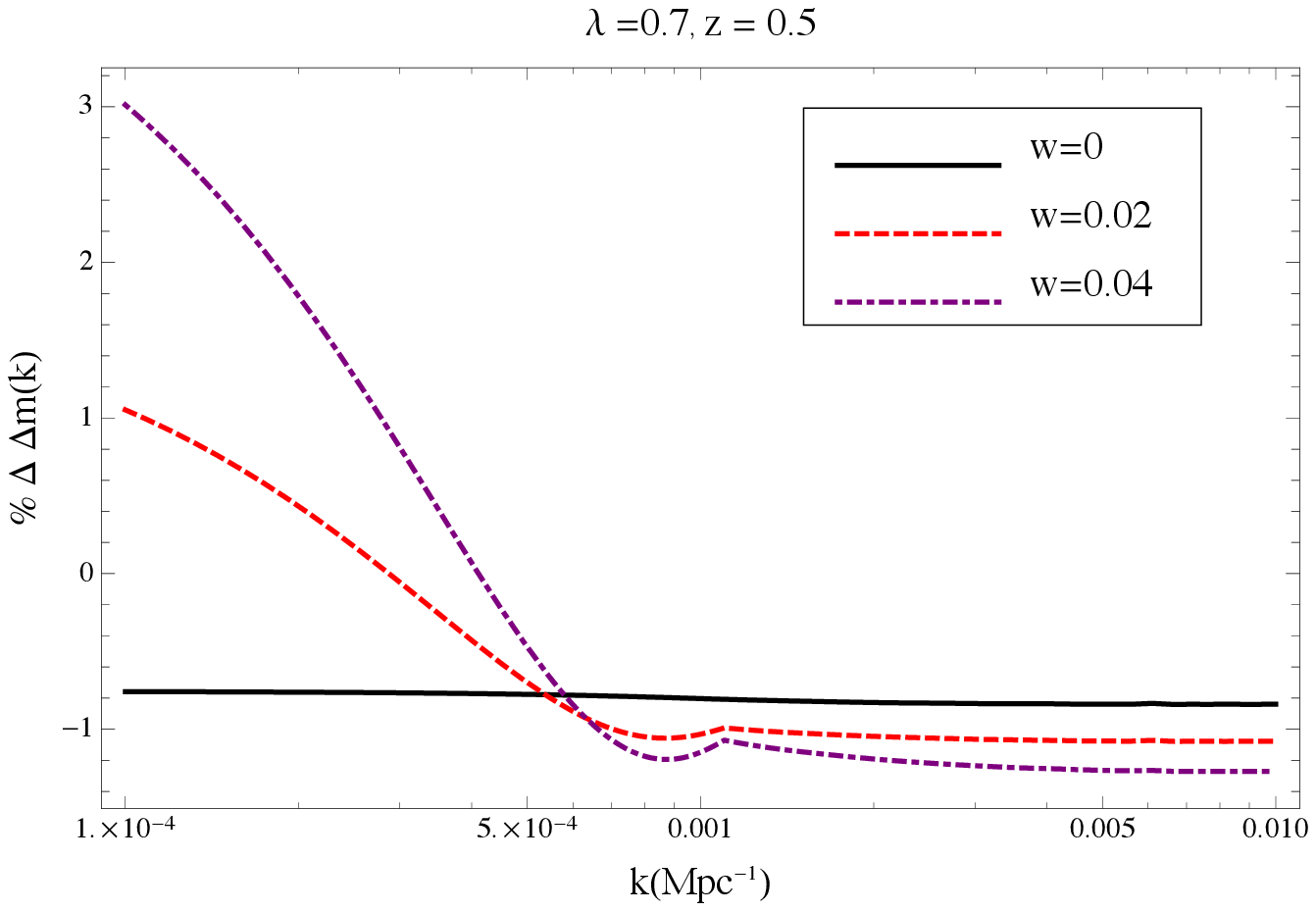,width=7.5 cm}\\
			\epsfig{file=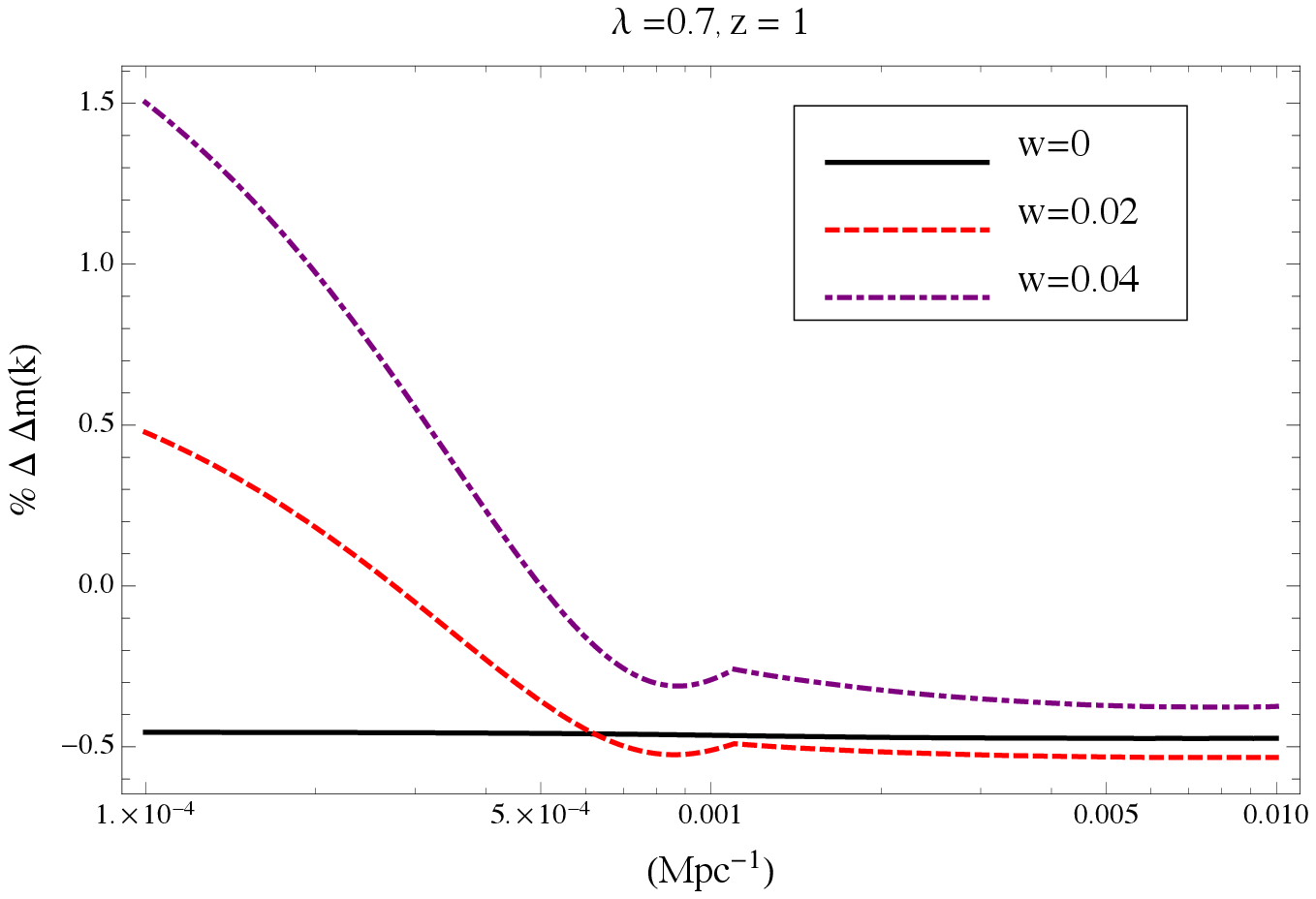,width=7.5 cm}
			\epsfig{file=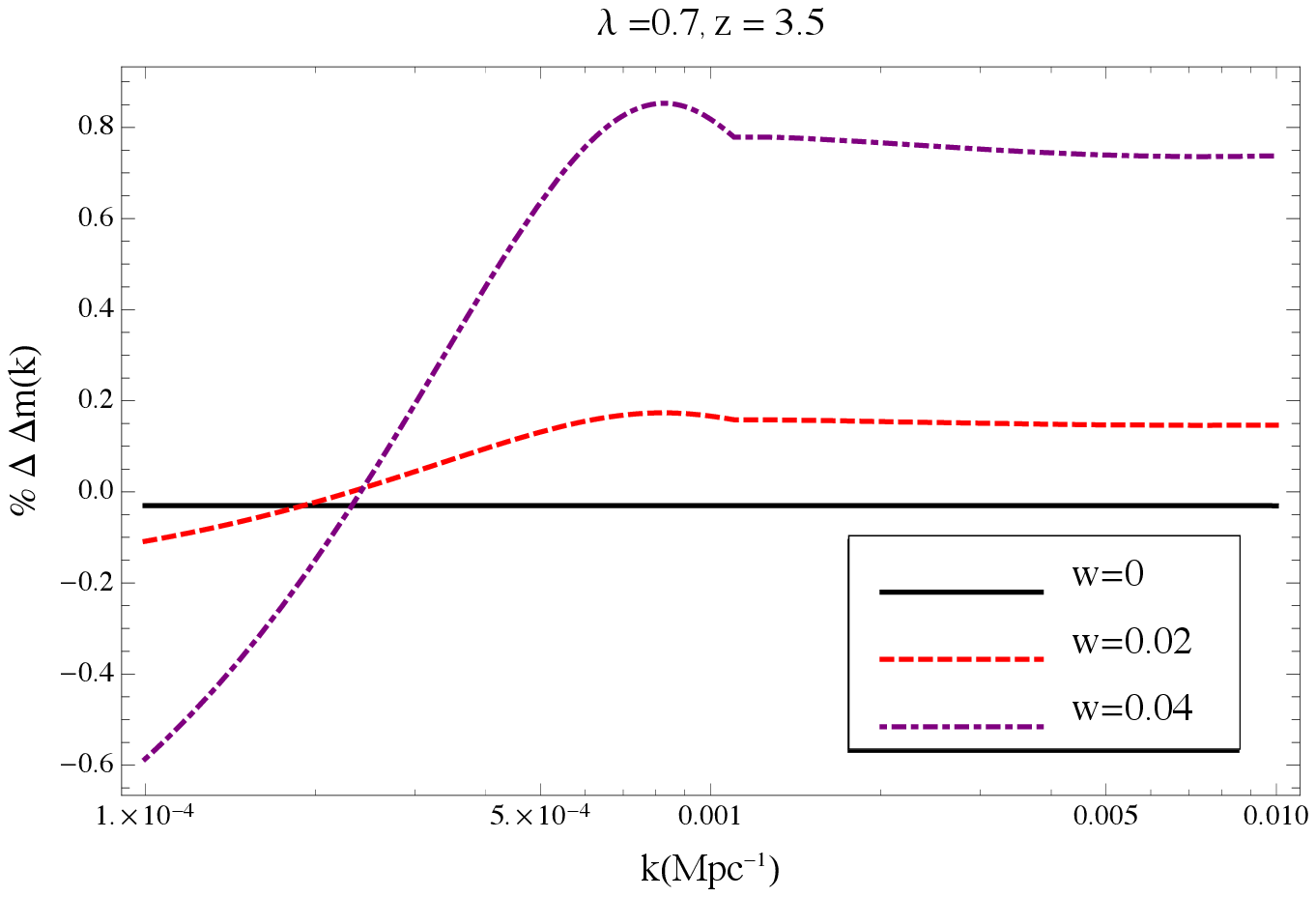,width=7.5 cm}
		\end{tabular}
		\caption{Percentage diversion of  comoving density contrast $ \Delta_{m} $ in IQ model as compared to  $ \Lambda$CDM model .
		}
	\end{figure*}
\end{center}
\begin{center}
	\begin{figure*}
		\begin{tabular}{c@{\quad}c}
			\epsfig{file=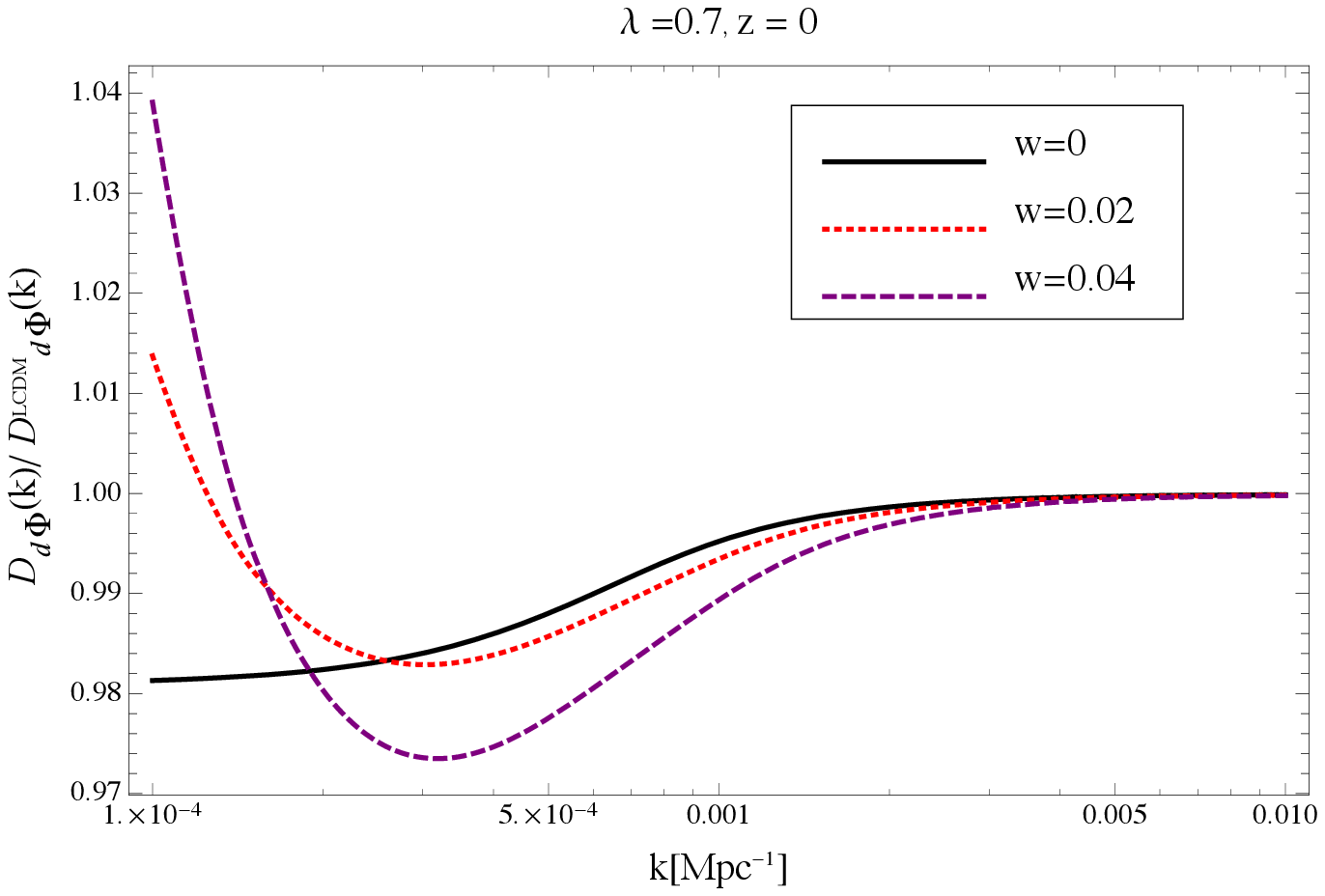,width=7.5 cm}
			\epsfig{file=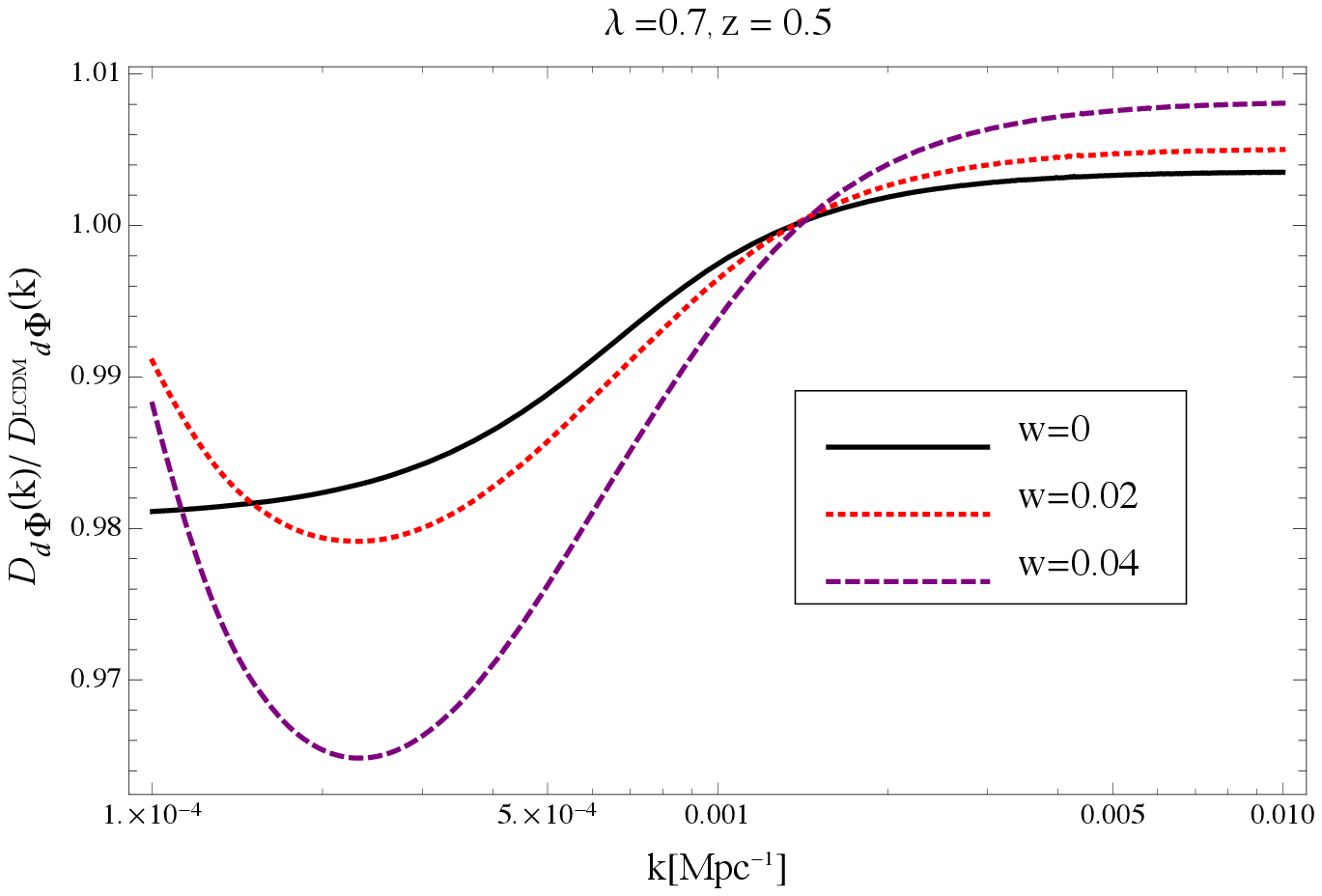,width=7.5 cm}\\
			\epsfig{file=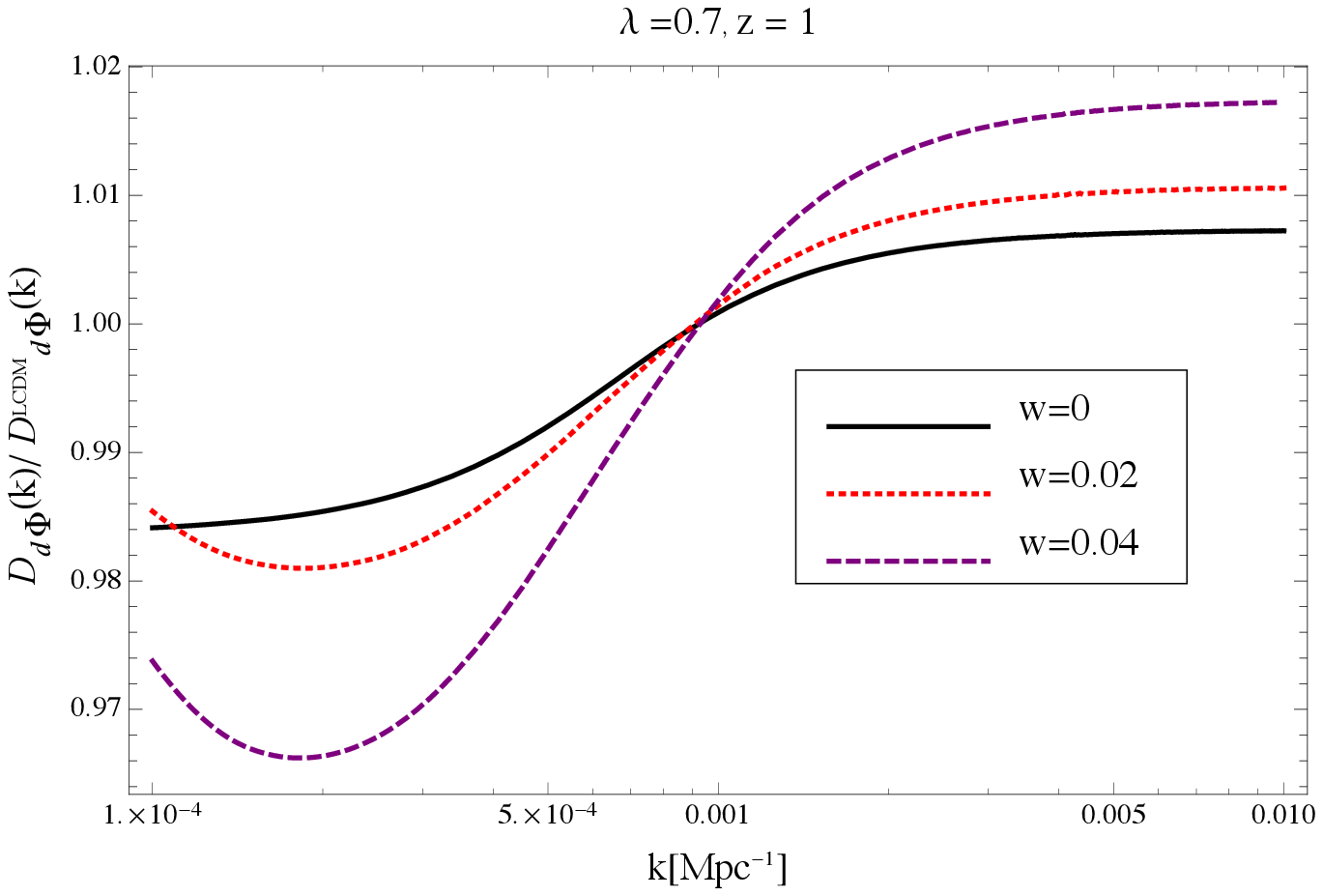,width=7.5 cm}
			\epsfig{file=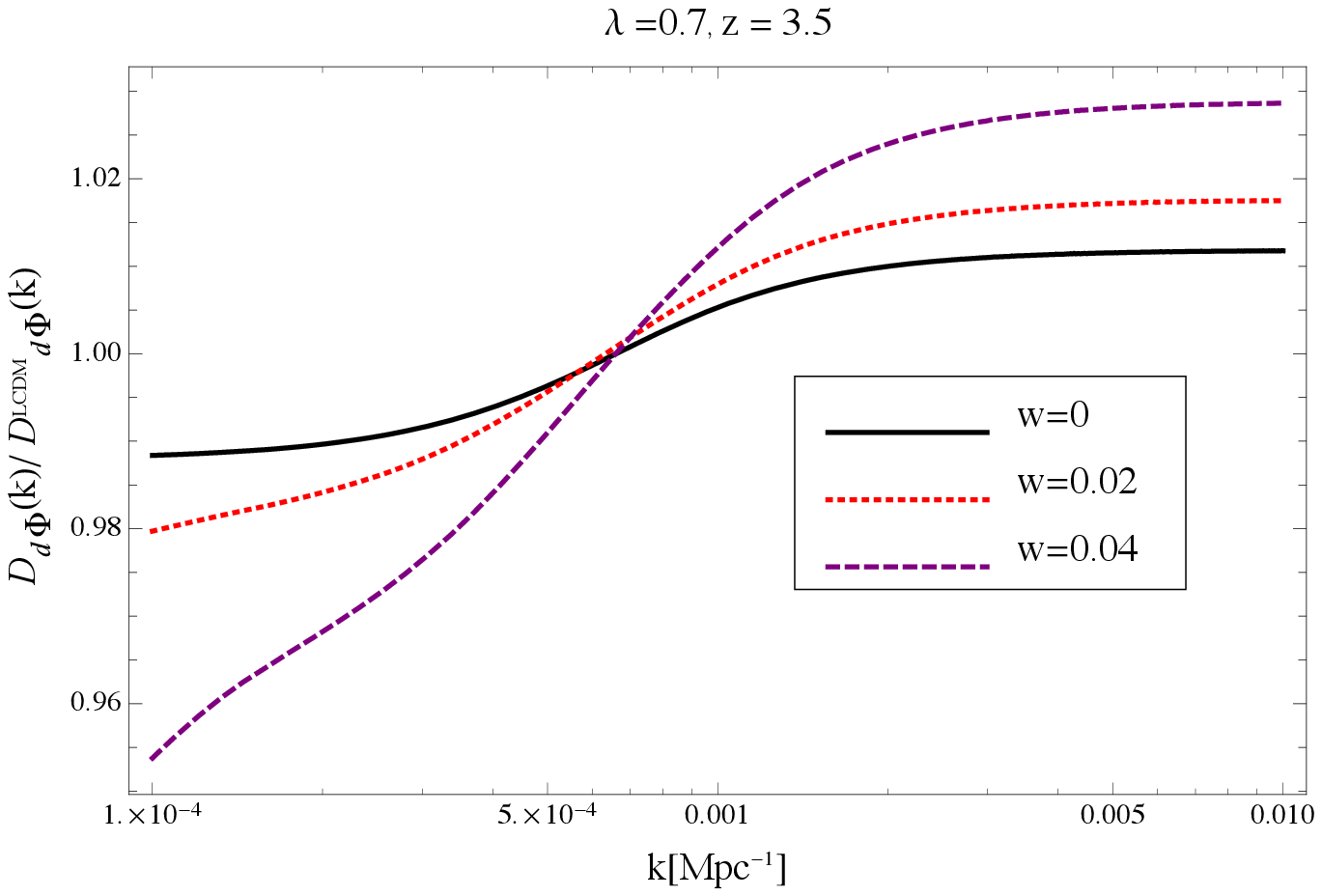,width=7.5 cm}
		\end{tabular}
		\caption{Ratio of $D_{d\Phi}(k,z)$ IQ model and $\Lambda$CDM defined in equation (28) .
		}
	\end{figure*}
\end{center}
\begin{center}
	\begin{figure}
		\begin{tabular}{c@{\quad}c}
			\epsfig{file=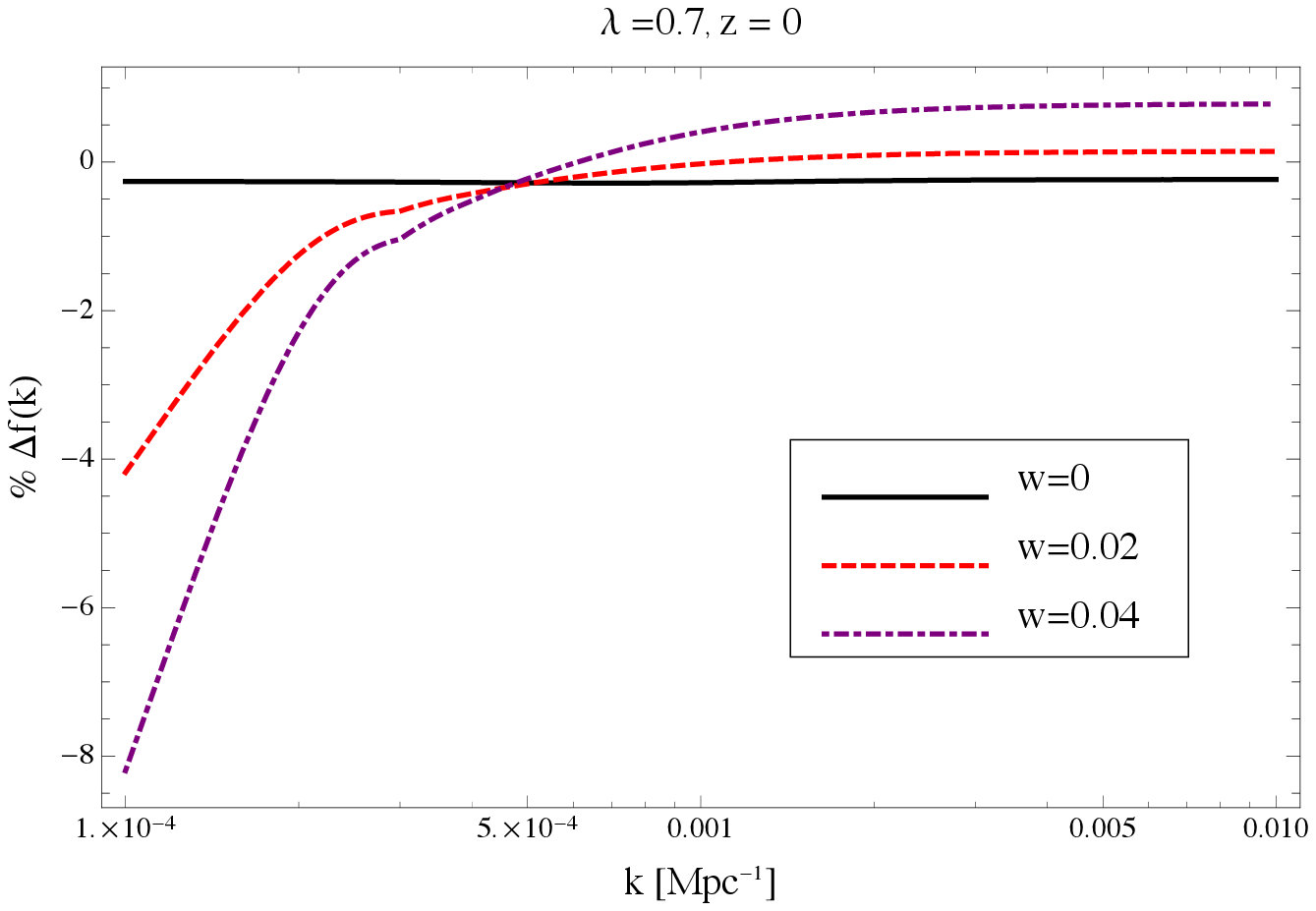,width=7.5 cm}
			\epsfig{file=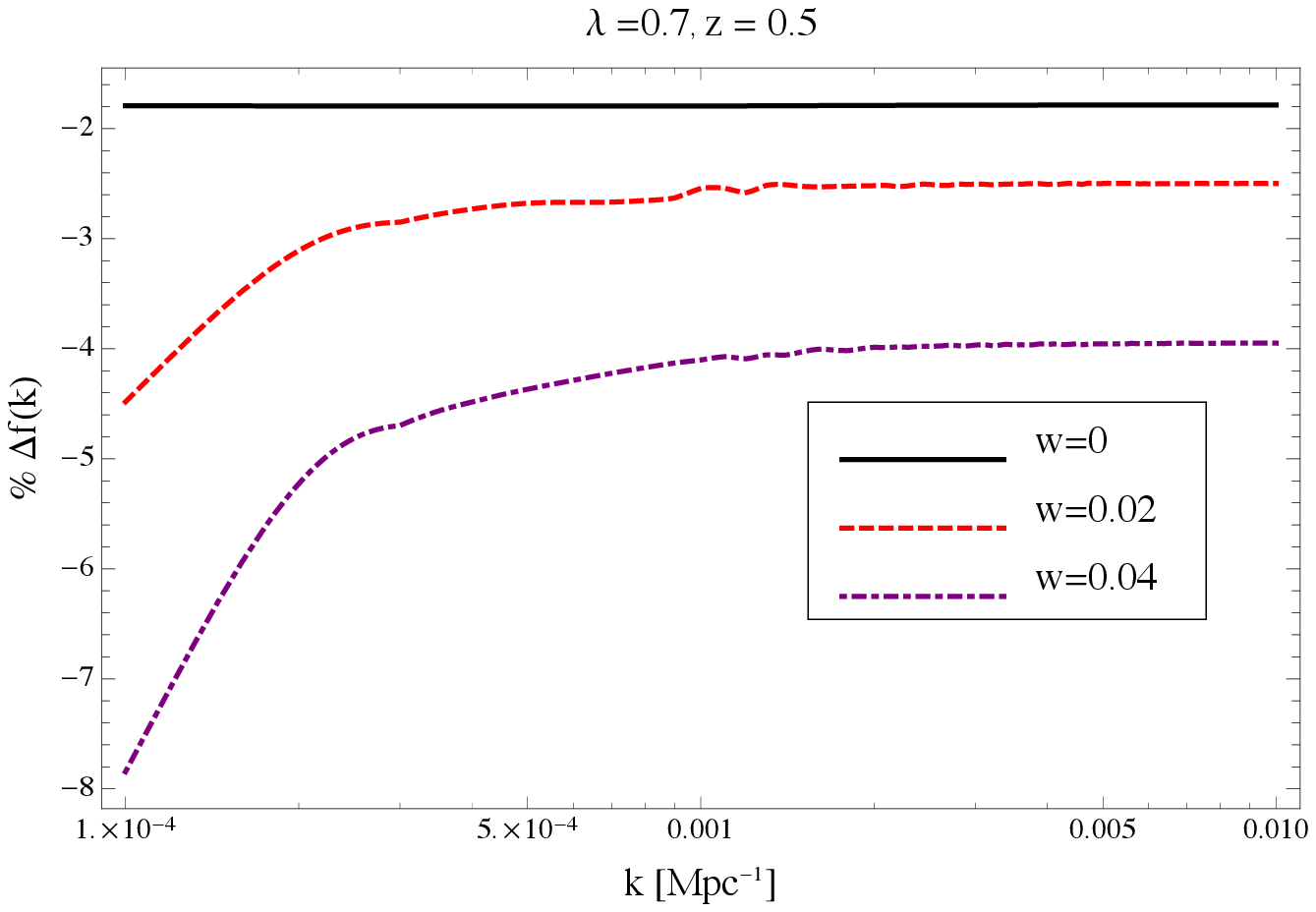,width=7.5 cm}\\
			\epsfig{file=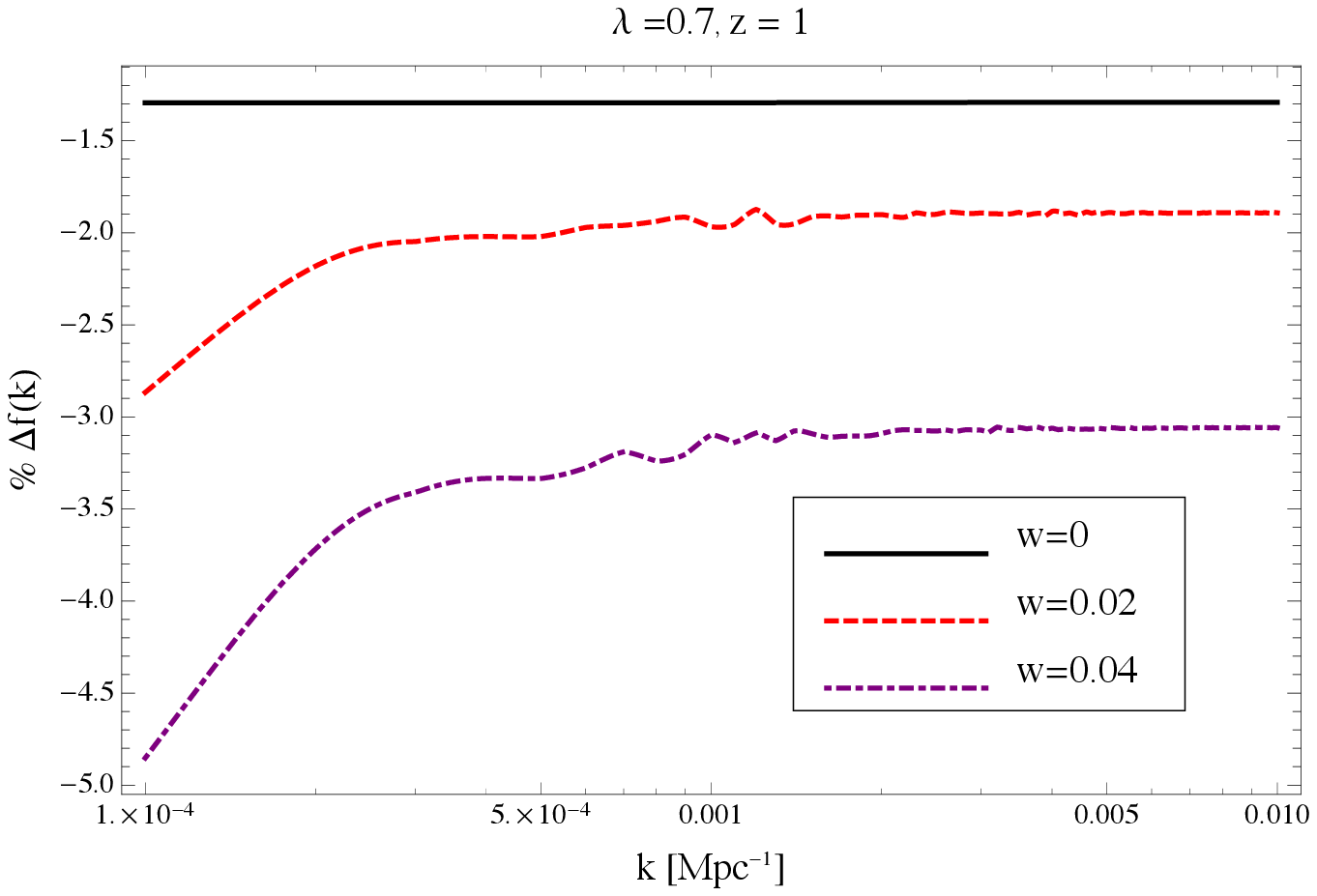,width=7.5 cm}
			\epsfig{file=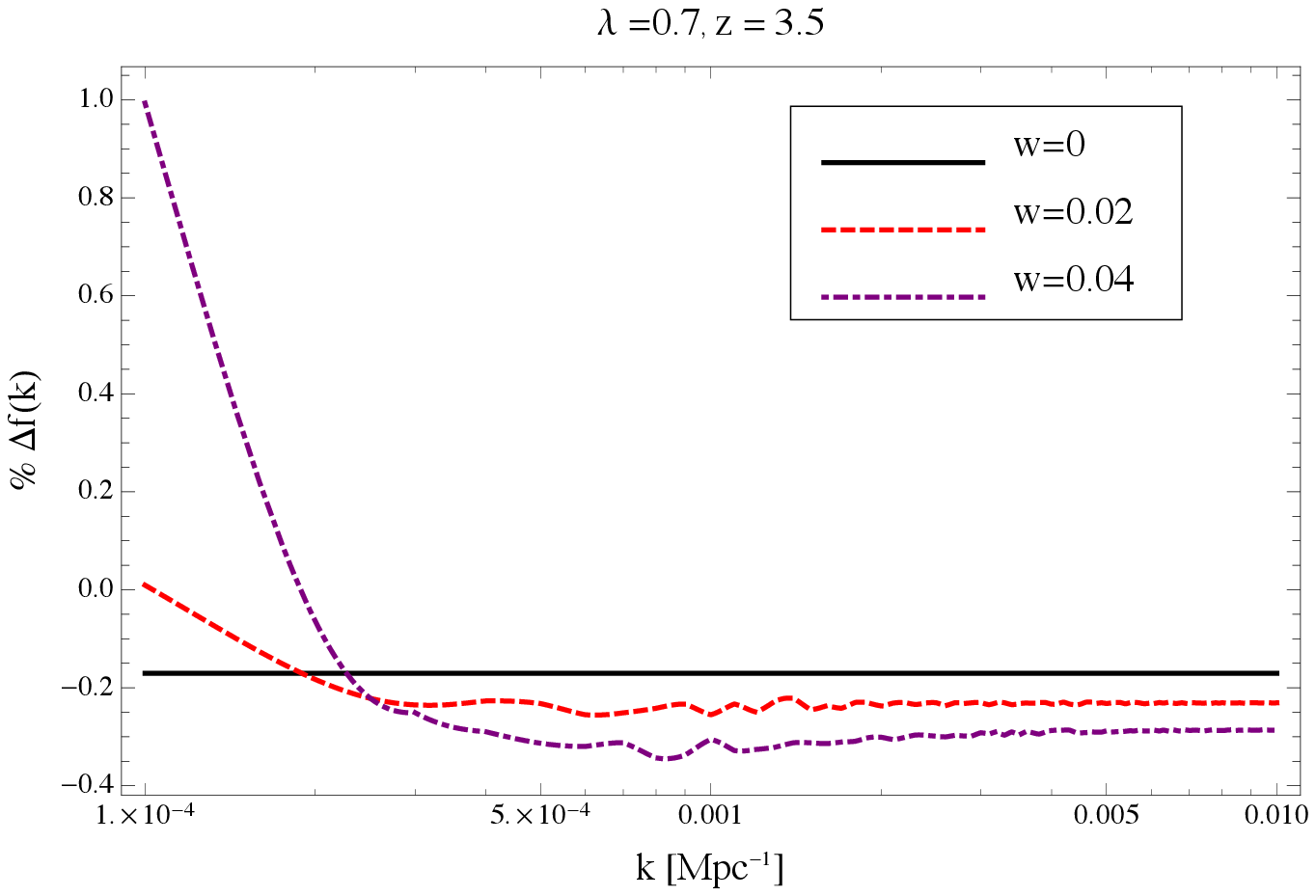,width=7.5 cm}
		\end{tabular}
		\caption{Percentage deviation in $f$ from $ \Lambda$CDM model.}
	\end{figure}
\end{center}
\subsection{Behaviour of cosmological parameters}

Using aforementioned initial conditions we solve perturbed equation (19) and (20 ) and study various pertubation  parameters, for different sets of interacting parameter W.

Figure 4 displays deviation in gravitational potential from LCDM model for different values of interacting parameter. For low redshift on sub-hubble scale there is a suppression from LCDM , which is due to different background evolutions as there is no contribution from DE perturbation at smaller scales. However on larger scale the enhancement in $\Phi$ is due to contribution from dark energy perturbtion . On increasing the value of interacting parameter  W there is overall enhancement on subhubble and super hubble scale respectively.This is due to transfer of energy and momenutm from DM to DE in both background as well as perturbed Universe. 

Figure 5 displays  variation in comoving matter density contrast $\Delta_{m}$. For non-interacting case variation is very small and is almost scale independent. At low redshift ($z=0,0.5,1$) on large scale  an increasing in  interaction (W) results in an enhancement in $\Delta_{m}$ wrt LCD ($ 0-8.5\%$) which decrease with redshift. But on smaller scales $\Delta_{m}$ is slightly suppressed less than $ 1\%$ . At redshift z=3.5 on large scale  there is slight suppression wrt lcdm less than$ 1\%$ and on small scale slight enhancement  less than$ 1\%$.

Figure 6 displays variation of f defined in equation (31) which is related to velocity perturbation and hence redshift space distortion. For non interactig case just like $\Delta_{m}$ variation is small and scale independent. On adding interaction its behaviour is exactly  opposite to that of matter density contrast $\Delta_{m}$. On large scale with increasing interaction a suppression can be seen at red shift 0,0.5 and 1 ($4-8\%$). But at higher redshift at $z=3.5$ there is slight enhancement less than $ 1\%$ . 

Hence interaction effect growth of structure at all scales.But this effect is smaller at higher redshift.

\section{  Influence of Interaction  on  power spectrums}
 
The growth of large scale structure in the Universe is ascertain by matter power spectrum. Forthcoming  surveys of galaxies can probe distribution of dark matter on large scales .These surveys  can provide strong bound on dark energy models including interaction in dark sector. We must incorporate  observed galaxy distribution effects like redshift space distortion , GR effects like weak lensing convergence , SW, ISW, time delay effect in our analysis to recognize potentiality of these surveys \cite{Yoo2010ni,Challinor2011bk,Jeong2011as,Bertacca2012tp,Hu2001yq}. Certain astrophysical processes such as gas cooling, star formation and feedback from supernovae, in conjunction with the gravitational effect of the dark matter has a bearing upon formation of galaxy. This can further cause a contrast between the spatial distribution of baryons and dark matter. The association between the spatial distribution of galaxies and ubiquitous dark matter must be understood to employ galaxies as cosmolgical probes known as galaxy bias.
Matter power spectrum can be related to galaxies distribution through  bias $b$ defined as \cite{Challinor2011bk,Bruni2011ta,Jeong2011as}

\beq\label{bias}
\Delta_{\rm g}(k,z) = b(z)\, \Delta_d(k,z).
\eeq
To study effect of interaction on matter and galaxy power spectrum we use prescription discussed by Duniya et al in \cite{L7}.

In a galaxy redshift survey, the observers measure the number of galaxies in direction n at redshift z.The  number overdensity of Galaxy () $\Delta^{obs}$) is shown as follows

\begin{equation}
\Delta^{obs} = \left[{b + f \mu^2} + \mathcal{A} (\frac{\mathcal{H}}{k})^2 +  i\mu\mathcal{B} (\frac{\mathcal{H}}{k})\right]\Delta_{m},
\end{equation}

 Here $\mu  = -\frac{\vec{n}.\vec{k}}{k}$ , $\vec{n}$ expressinng the direction of observation,  $f$ signifies  the redshift space distortion and b stands for galaxy bias. Variables A and B  which are considered in connection to GR corrections are delineated as follows:

\begin{equation}
A=(3-b_{e})f-\frac{3\Omega_{d0}H_{0}^{2}}{2{\cal H}^{2} D_{d}}(4Q-b_{e}-1-(1+z)\frac{{\cal H}^{'}}{\cal H}+2\frac{(1-Q)}{r{\cal H}} + \frac{1}{D_{\Phi}(1+z)^{2}}(-(1+z)^{2}(D_{\Phi}+(1+z)D_{\Phi}^{'})D_{\Phi}(1+z)
\end{equation}

\begin{equation}
 B=[b_{e}-2Q+(1+z)\frac{{\cal H}^{'}}{\cal H}-2\frac{(1-Q)}{r{\cal H}}-w\sqrt{6}x(1-\frac{Dv_{\phi}}{Dv_{d}})]f
\end{equation}

Full general relativistic power spectrum that includes all GR effects \cite{L7}.
\\
\begin{equation}
P_{g}^{obs}(k,z)=((b+f\mu^{2})^{2}+2(b+f\mu^{2})\frac{A{\cal {H}}^{2}}{k^{2}}+\frac{A^{2}{\cal {H}}^{4}}{k^{4}}+\mu^{2}\frac{B^{2}{\cal {H}}^{4}}{k^{4}})P(k,z)
\end{equation}

Matter power spectrum with kaiser term is given by
\begin{equation}
P_{k}(z,k) = \big[b(z) + f(z,k)\mu^2\big]P(z,k)
\end{equation}
Here prime is derivative wrt redshift.\\
A constant comoving galaxy number density is presumed hence $b_e =0$ , galaxy bias $b=1$ and the magnification bias ${Q} = 1,$. Here x is defined in equation (4). A is in connection with  peculiar velocity potential and the gravitation potential, on the other hand  B is connected to the Doppler effect. The latter comprises an interaction  explicitly. However no momentum is transported  in the Dark energy rest frame which leads to last term in equation (35) would be zero \cite{L7} .
\begin{center}
	\begin{figure*}
		\begin{tabular}{c@{\quad}c}
			\epsfig{file=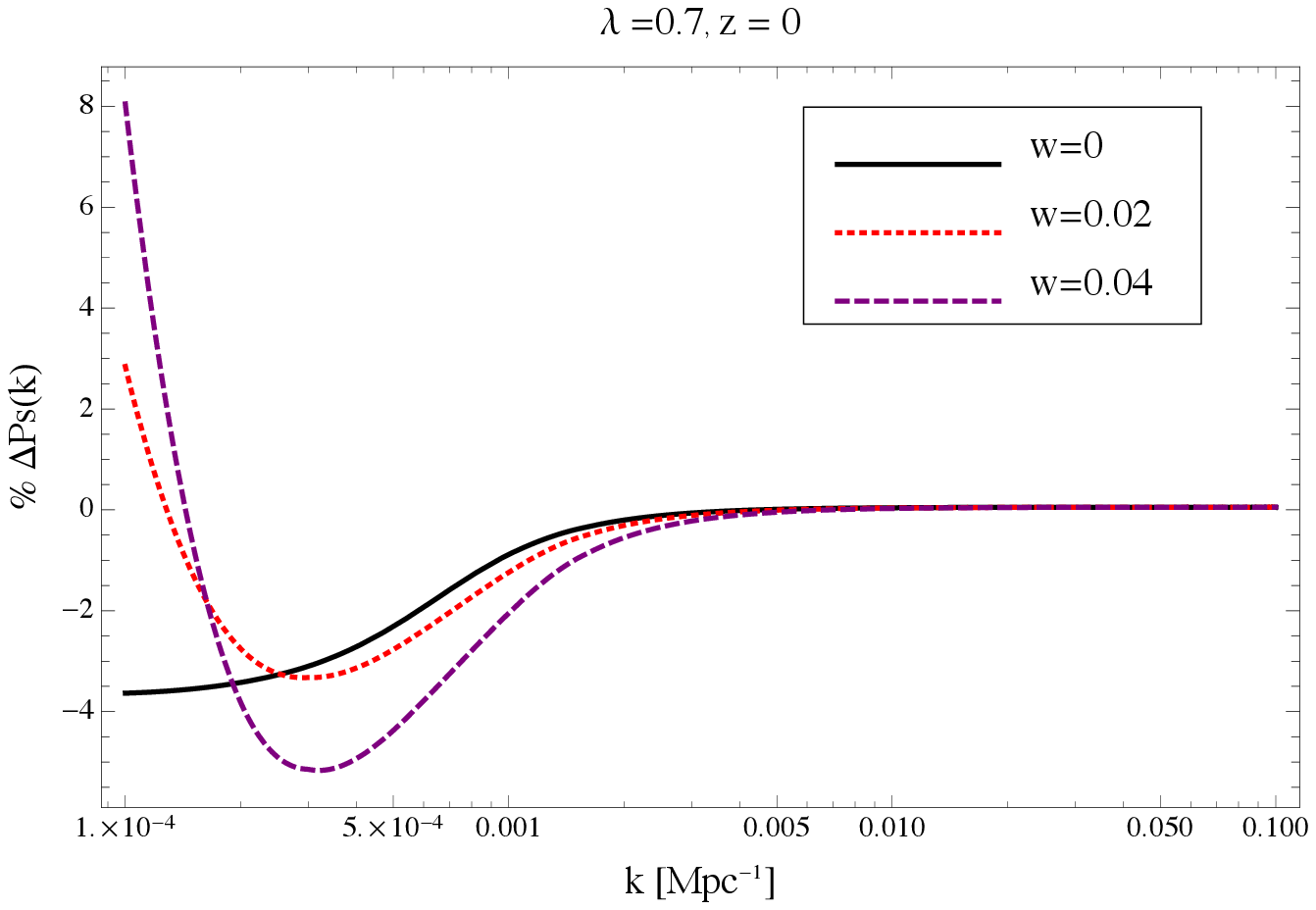,width=5.5 cm}
			\epsfig{file=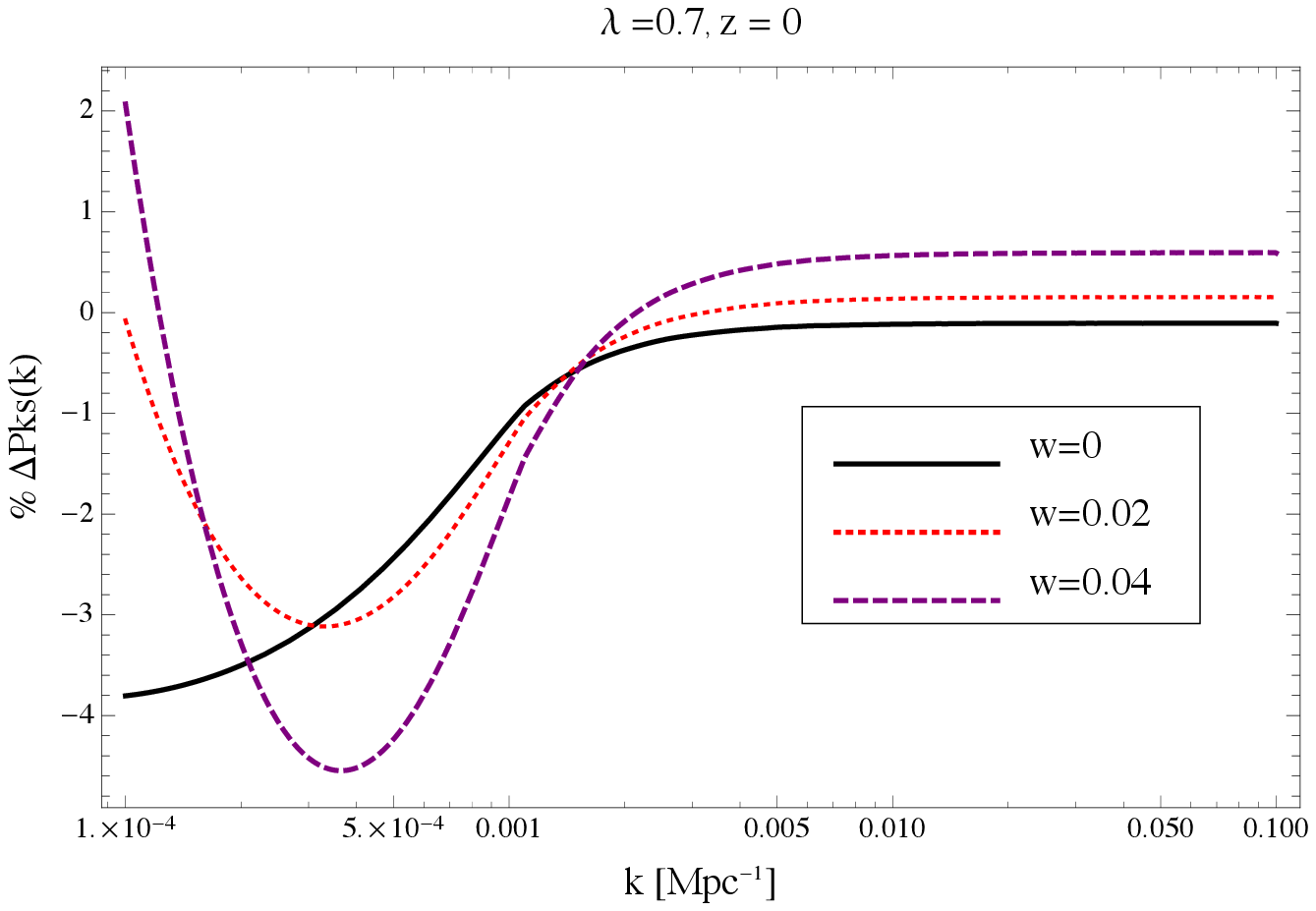,width=5.5 cm}
			\epsfig{file=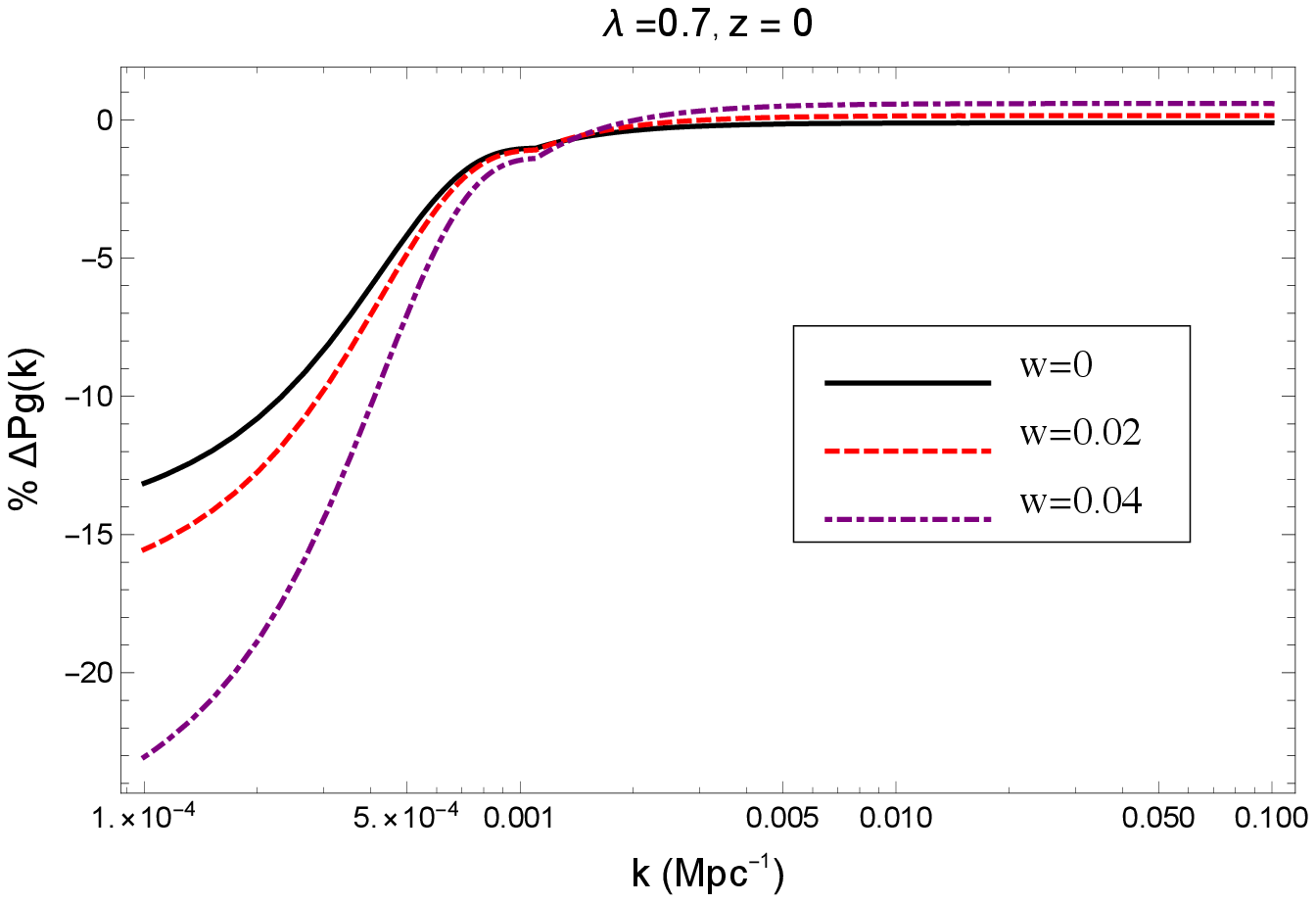,width=5.5 cm}\\
			\epsfig{file=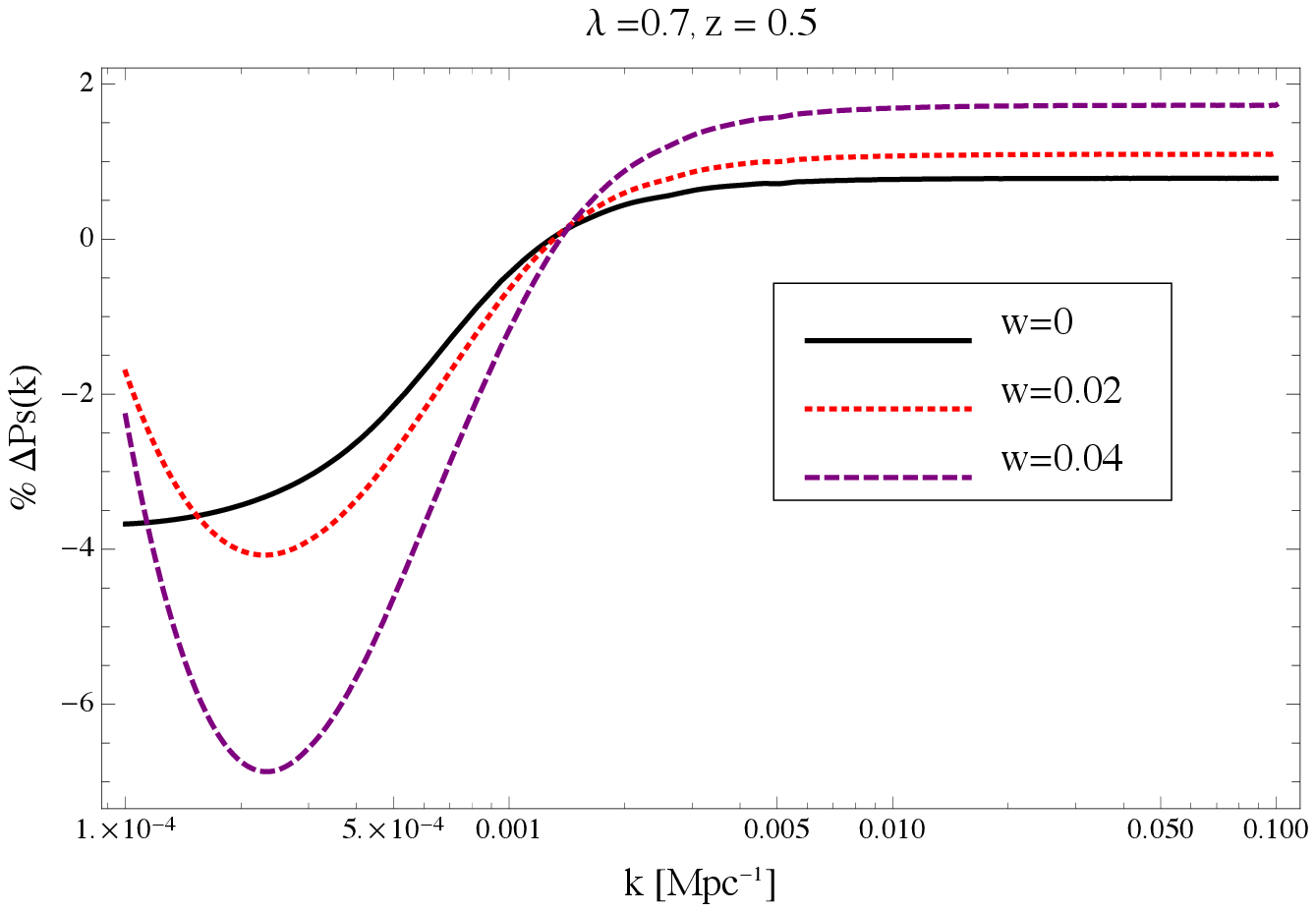,width=5.5 cm}
			\epsfig{file=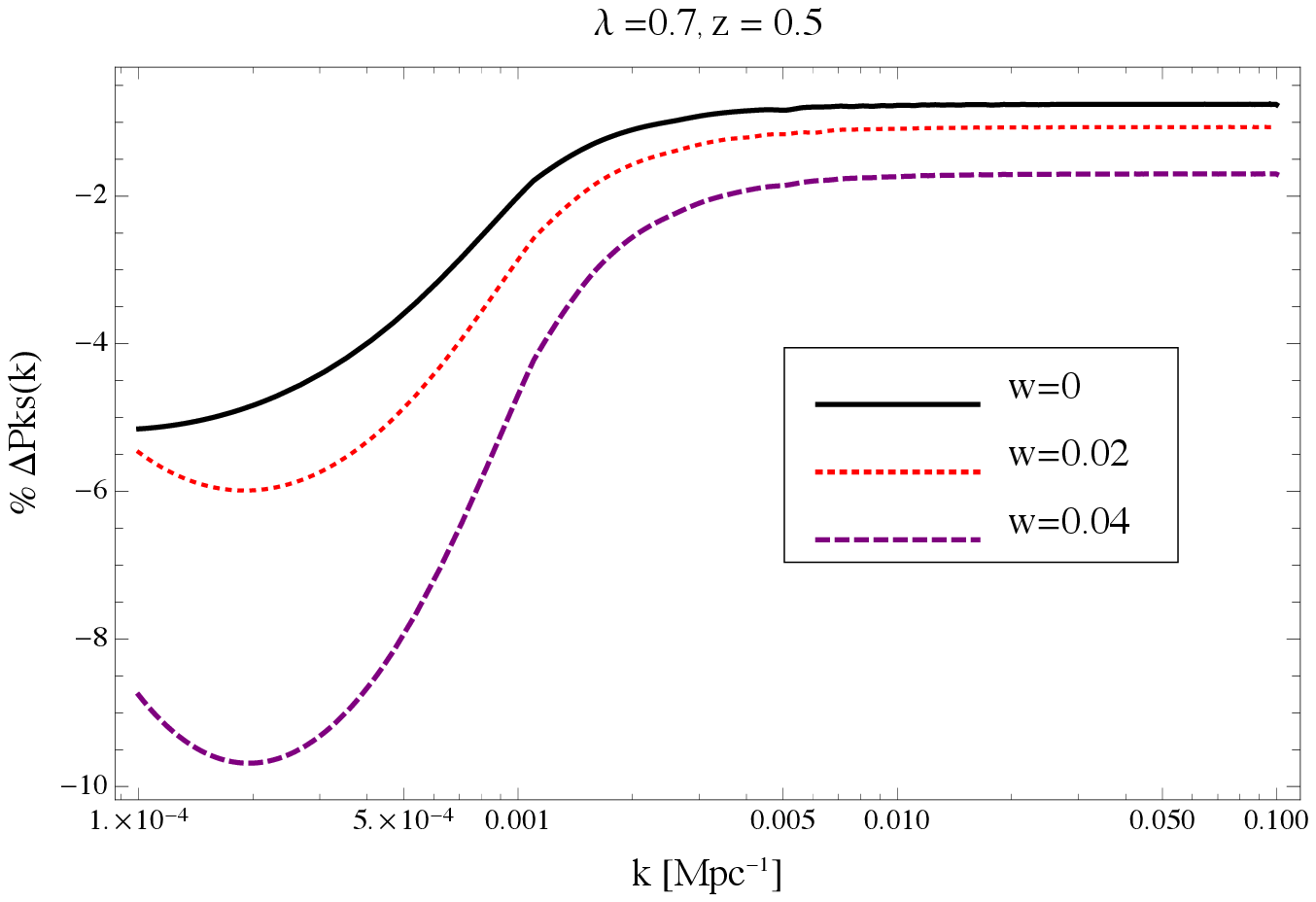,width=5.5 cm}
			\epsfig{file=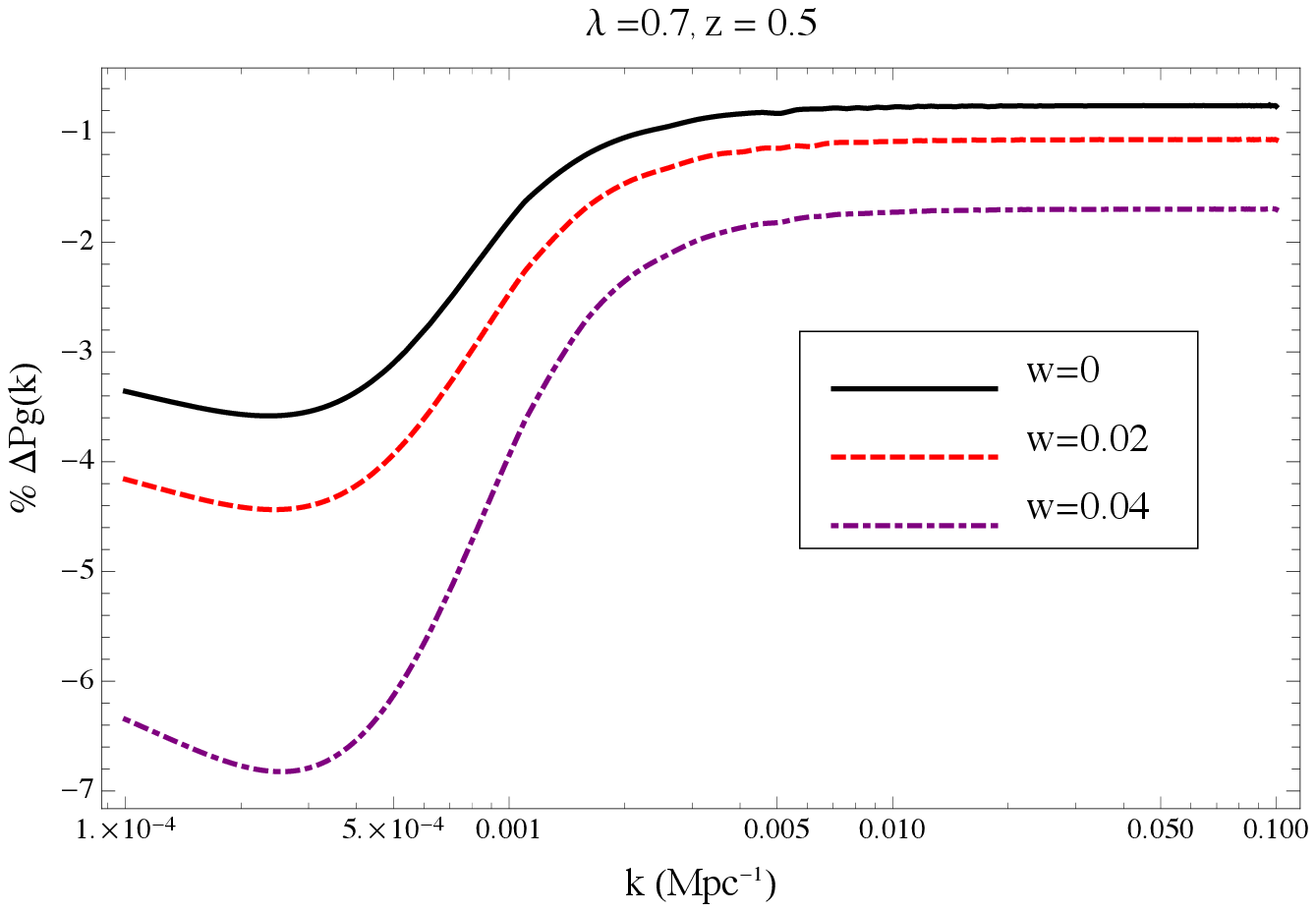,width=5.5 cm}\\
			\epsfig{file=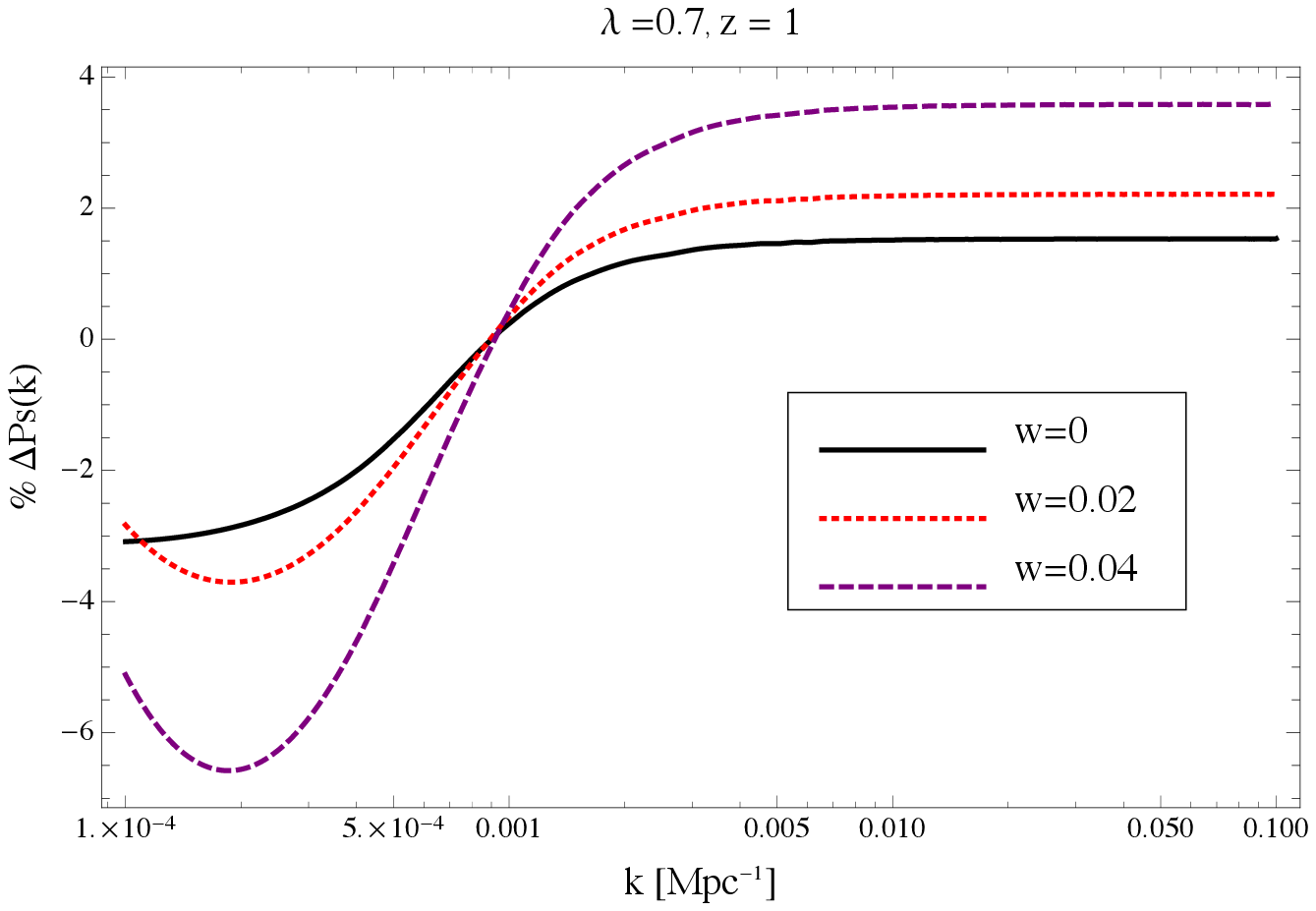,width=5.5 cm}
			\epsfig{file=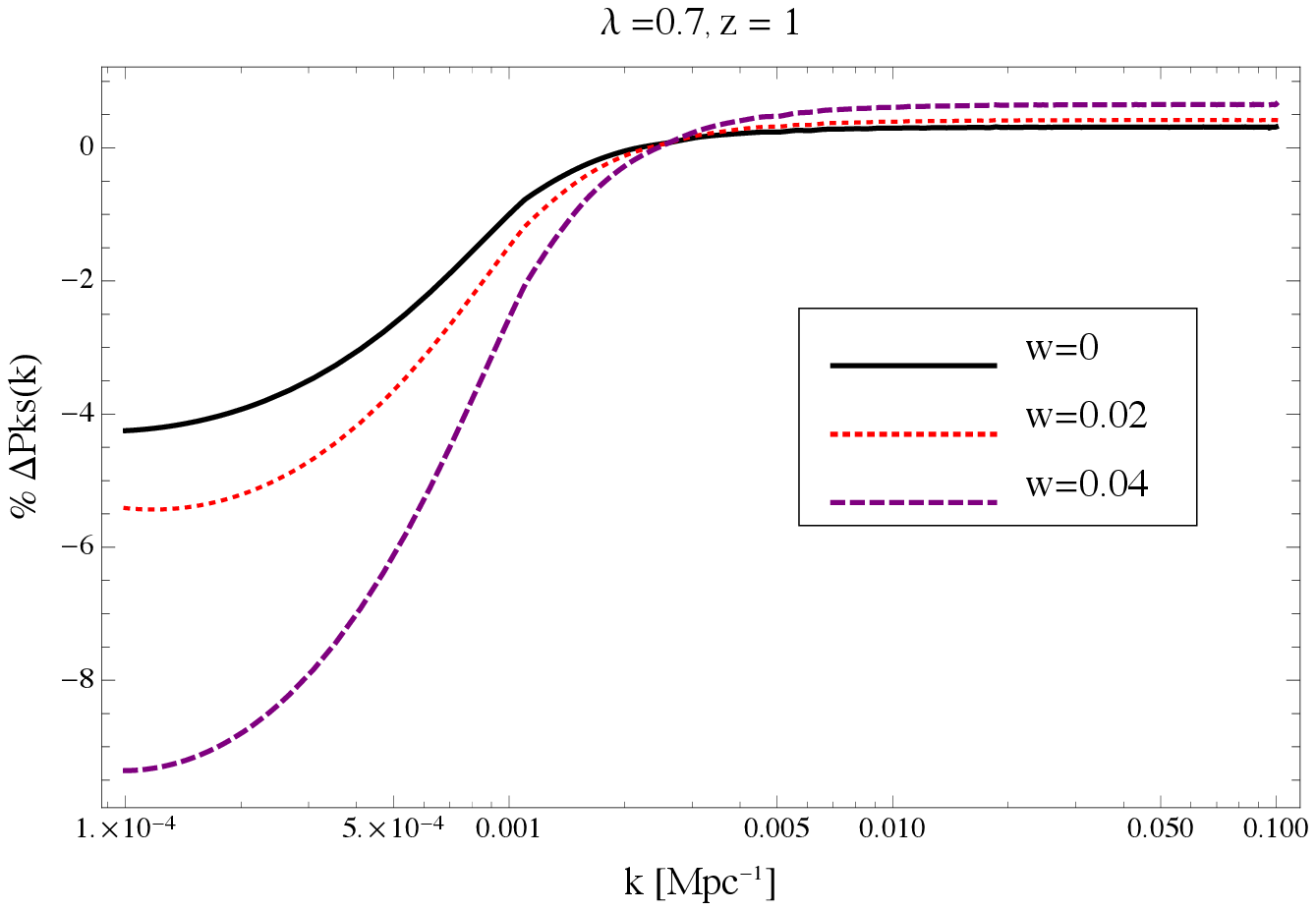,width=5.5 cm}
			\epsfig{file=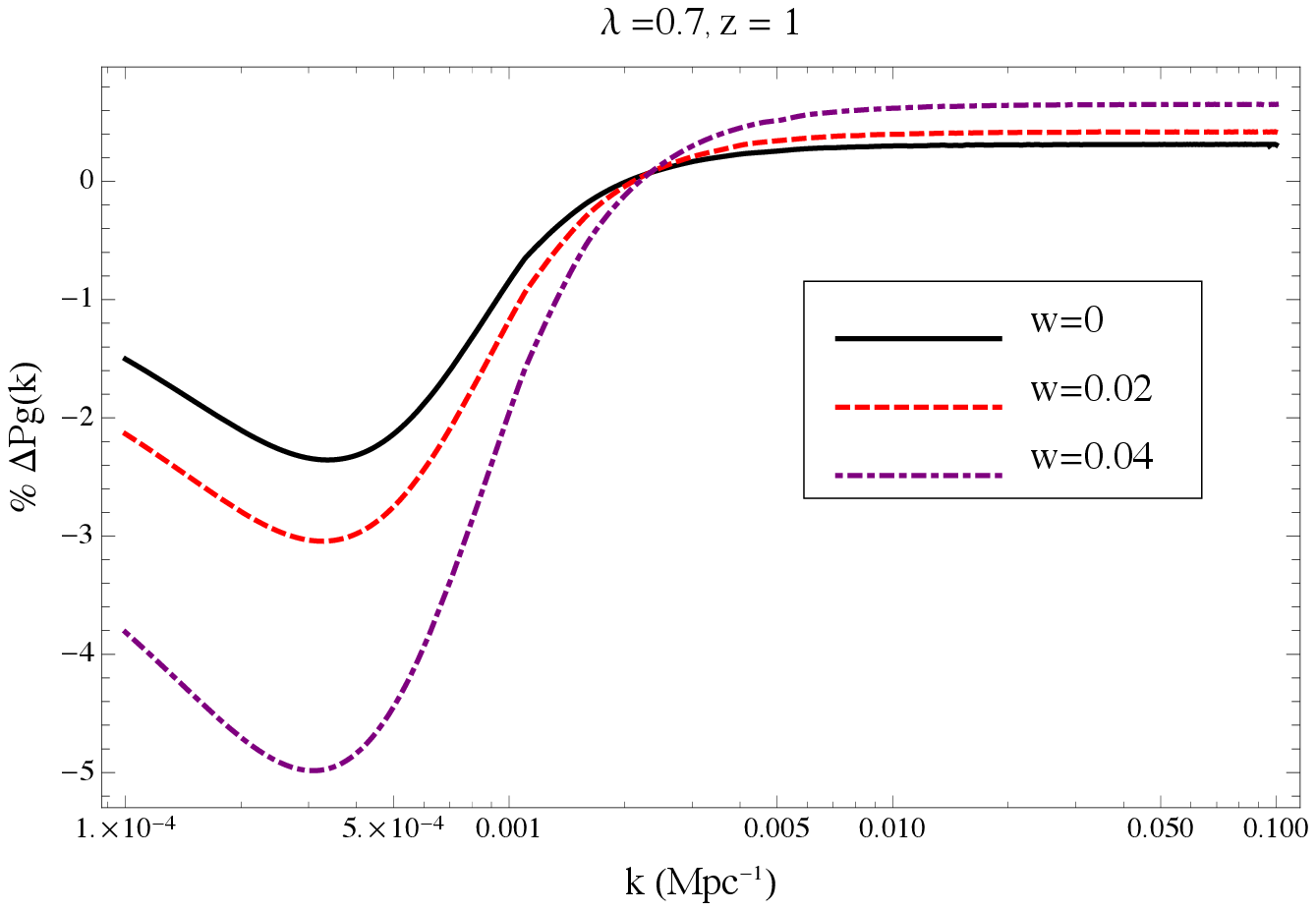,width=5.5 cm}\\
			\epsfig{file=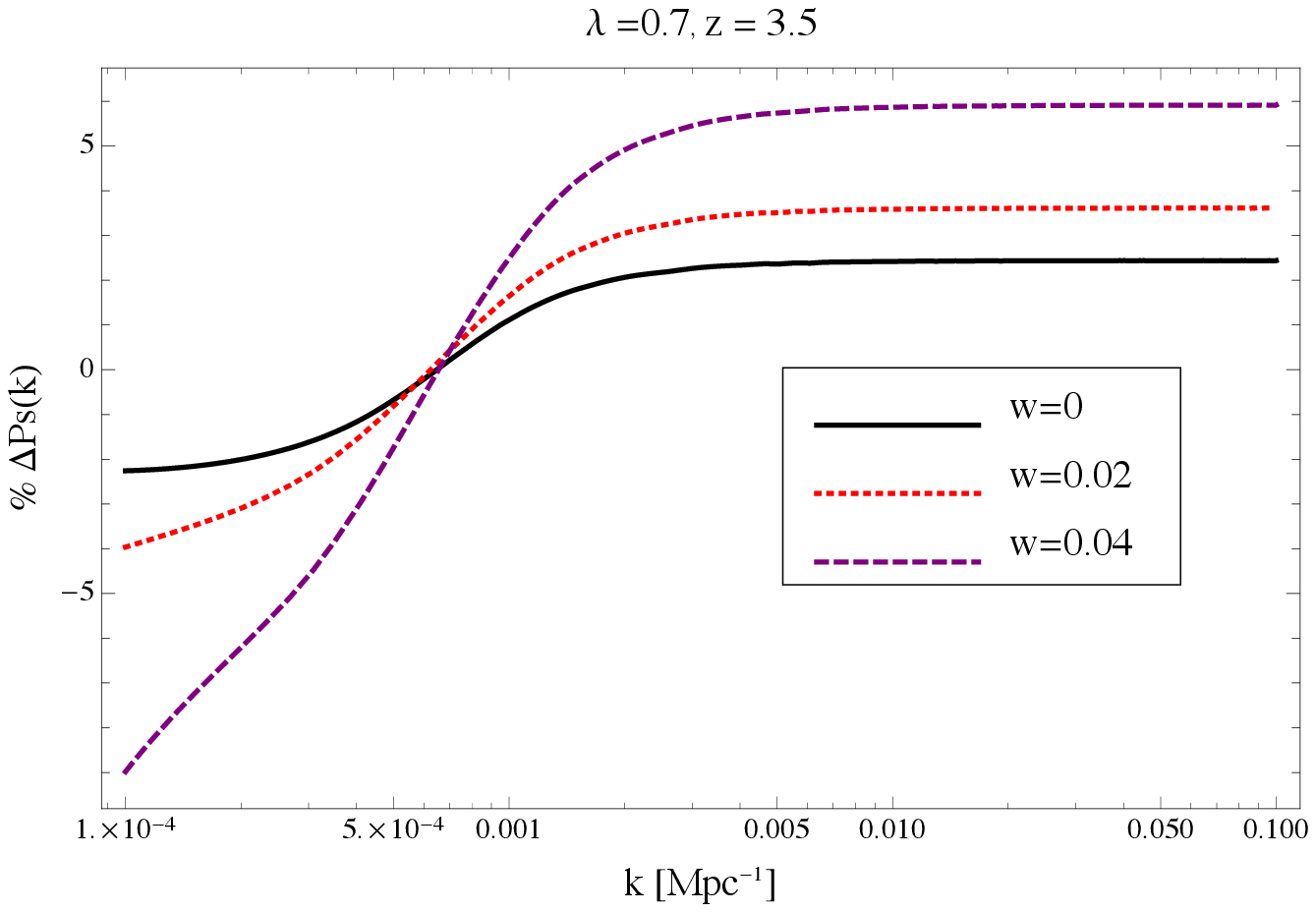,width=5.5 cm}
			\epsfig{file=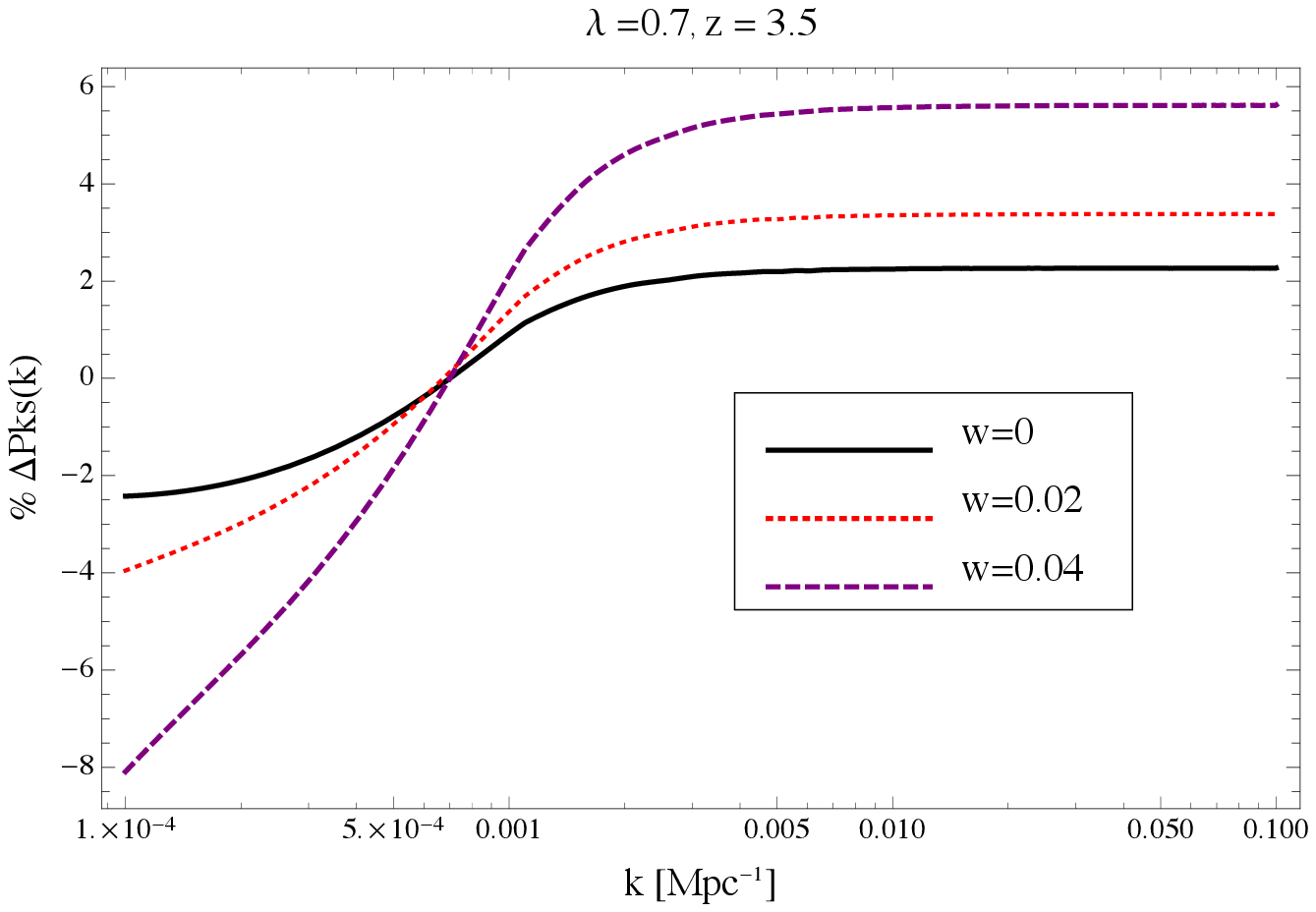,width=5.5 cm}
			\epsfig{file=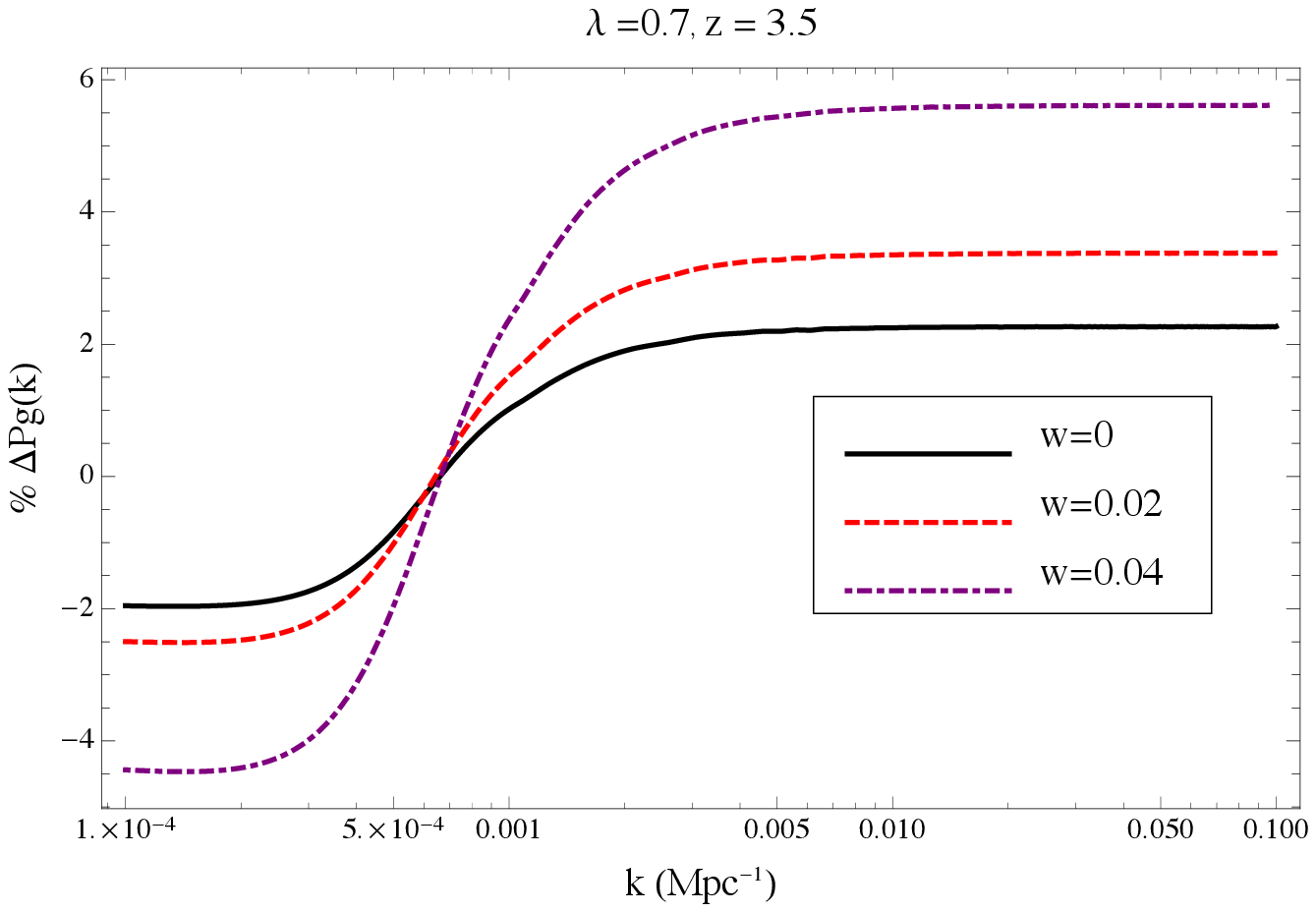,width=5.5 cm}\\
		\end{tabular}
		\caption{Percentage of diversion in $ P (k) $ in IQ model as compared to  $ \Lambda$CDM model for different values of interacting parameter W as a function of $K$ for different redshifts. First column is standard matter power spectra $P$ which is specified by eqn. (30), Second column present Kaiser power spectra $P_{k}$ specified by eqn (37), third column portrays  galaxy power spectrum specified by eq (36).
		}
	\end{figure*}
\end{center}
Figure 8 (left column) shows percentage change in standard matter power spectrum wrt $\Lambda$CDM. It depends on ratio $\frac{\Delta_{d}}{\Phi}$ which is shown in figure (6).  At $z=0$ and on large scale percentage change for non-interacting case ($W=0$) is -ve which shows suppression in power.But on addding interaction  ($2-8\%$) enhancement is observed.Which shows transfer of energy from scalar field to dark matter , rate of transfer increase with increase in interaction parameter W.On smaller scale all models converge to$\Lambda$CDM due to our Normalization.On larger scale and higher red shift dark energy gives negative contribution as compared to $z=0$ .Thus suppression is seen wrt $\Lambda$CDM. This suppression contribution increases with redshift.But on smaller scales slight   enhancement of power can be seen which  is due to difference in background evolution.

Figure 8 (middle column) displays percentage change in matter power spectrum with kaiser term wrt $\Lambda$CDM. At $z=0$ on large scale enhancement in kaiser spectrum is  ($0-2\%$) which is less than standard matter power spectrum discussed earlier. Which is due to contribution from kaiser term eq (37) that give negative contribution to power spectrum. It  depends on parameter f eq (33) . Behaviour of f is shown shown in fig 7 which is suppressed wrt $\Lambda$CDM. On higher redshift suppression is more as compared to standard power spectrum . But on smaller scale all models converge to$\Lambda$CDM for all redshift.
 
 Figure 8 (right column) displays percentage change in galaxy power spectrum wrt  $\Lambda$CDM. At $z=0$ for large scales percentage change is negative ($12-23\%$) which shows suppression wrt $\Lambda$CDM. On increasing interaction this suppression increases and it decreases with redhsift. The reason of suppression is general relativistic terms A and B  in equation (34) which give negative contribution. On small scales and low redshift percentage change is small but on high red shift $z=3.5$ percentage change is positive ($2-6\%$).The reason for this is weaker dark energy effect at higher z but GR effects are  comparative stronger even on small scales.  
 
  \section{Conclusion}
  
 We have generalised non-interacting quientessence model as discussed by Bikas et al \cite{Bikash} for interacting scenerio.
 We studied effect of interaction between Quientessence and dark matter on both background and perturbed universe. At background level interaction effect bacground energy density and  hubble parameter .Suppression in dark matter density $\Omega_{d}$ wrt $\Lambda$CDM is  ($0-8\%$) . While enhancement in hubble is ($0-8\%$). Thus interaction makes dark energy effect stronger.
 
 A careful analysis of growth of strucure characterises that interaction effects matter, kaiser and galaxy power spectrum on sub and super horizon scales. The former is because of dark energy perturbation and GR effect and the latter is the result of background evolution.
 
 We find that Standard matter power spectrum at  $z= 0$ is enhanced wrt $\Lambda$CDM on large scales and on small scale no deviation due to our normalization. This enhancement on large scale increases with increase in interaction. At higher redshift matter power spectrum suppressed wrt $\Lambda$CDM. This suppression also increases with interaction.
 On adding kaiser redshift distortion term enhancement  is less as compared to standard matter power spectrum. Thus it can be concluded that that kaisr term gives negative contribution to power spectrum.Which suppress power on large scale.On higher redshift it is further suppressed .
 
 We also found that Galaxy power spectrum is suppressed ($12-24\%$) wrt $\Lambda$CDM on large scales .This suppression increases with interaction. But on higher redshift effect of dark energy is weaker hence suppression decreases with increase in z. But on small scales  enhancement is observed ($2-6\%$) which is due to difference in background evolution and GR effects .Thus on higher redshifts interaction effects galaxy power spectrum even on smaller scale.
 This deviation can be probed by future surveys like SKA.


\begin{thebibliography}{99}%
	
	\bibitem{per}
	Perlmutter, S. {\it et al.} Astrophys.J. 483 (1997) 565 astro-ph/9608192 FERMILAB-PUB-98-035-A, LBL-39291, LBNL-39291
	
	\bibitem{SN2}
	Perlmutter, S. {\it et al.} Astrophys. J. 517  (1999) 565 astro-ph/9812133.
	
	 \bibitem{SN}
	A.~G.~Riess {\it et al.} Astrophys. J.{{\bf 116}, 1009 (1998)}
  []{astro-ph/9805201}].
	
	   \bibitem{SN3}
	J.~L.~Tonry {\it et al.}  [Supernova Search Team Collaboration],
{Astrophys.\ J.\  {\bf 594}, 1 (2003)}
	[{astro-ph/0305008}].
	
	\bibitem{ade}
	P.A.R.Ade et al. [Planck Collaboration], Astron. Astrophys., {\bf 594}, A13 (2016).
	
	\bibitem{jla} 
	M.~Betoule {\it et al.}  [SDSS Collaboration],
	Astron.\ Astrophys.\  {\bf 568}, A22 (2014).
	
	  \bibitem{bao}
	A. Lauren et al., Mon.Not.Roy.Astron.Soc., {\bf 441}, 24 (2014);
	A. Lauren et al., Mon.Not.Roy.Astron.Soc., {\bf 427}, 3435 (2013);
	F. Beutler et al., Mon.Not.Roy.Astron.Soc., {\bf 416}, 3017 (2011);
	F. Beutler et al., Mon.Not.Roy.Astron.Soc., {\bf 423}, 3430 (2012);
	C. Blake et al., Mon.Not.Roy.Astron.Soc., {\bf 425}, 405 (2012).

	\bibitem{win}
    Weinberg S., Rev. Mod. Phys., {\bf 61}, 1(1989).
    
    \bibitem{hu}. 
       Huey G., Wandelt B. D., Phys.~Rev.~D. {\bf 74}, 083506 (2011).
       
       \bibitem{discrep}
       
       C.Heymans et al., Mon.Not.Roy.Astron.Soc., {\bf 432}, 2433 (2013);
       B.~A.~Reid et al., Mon.Not.Roy.Astron.Soc., {\bf 444}, 476 (al.2014);
       E. Abdalla et al.,  ``{\it Evidence for Interacting dark energy from BOSS}",  arXiv:1412.2777.
       
       \bibitem{riess} 
       A. Riess, et al., Astrophys. J, {\bf 826}, 56 ( 2016)
       
       \bibitem{kids}
       H. Hildebrandt et al.,  Mon.Not.Roy.Astron.Soc., {\bf 465}, 1454 (2017).
       
       \bibitem{valen}E. de Valentino et al., ``{\it Constraining Dark Energy Dynamics in Extended Parameter Space}",  arXiv: 1704 00762.
    
     \bibitem{de}
    V. Sahni and A.A. Starobinsky, Int. J. Mod. Phys. {\bf D9} 373 (2000);
    P.~J.~E. Peebles and B. Ratra, Rev. Mod. Phys. {\bf 75} 559 (2003);
    T. Padmanabhan, Phys. Rep. {\bf 380} 235 (2003);
    V. Sahni, astro-ph/0202076, astro-ph/0502032; 
    
    \bibitem{Ratra:1987rm} 
    B.~Ratra and P.~J.~E.~Peebles,
    Phys.\ Rev.\ D {\bf 37}, 3406 (1988).
    doi:10.1103/PhysRevD.37.3406
    
    \bibitem{Peebles:1987ek} 
    P.~J.~E.~Peebles and B.~Ratra,
    Astrophys.\ J.\  {\bf 325}, L17 (1988).
    doi:10.1086/185100
    
    
    \bibitem{Copeland:2006wr}
    E.~J.~Copeland, M.~Sami and S.~Tsujikawa,
    {Int.\ J.\ Mod.}\
    {Phys.\ D {\bf 15}, 1753 (2006)}
    [{hep-th/0603057}].
    
    \bibitem{Sahni:1999gb}
    V.~Sahni and A.~A.~Starobinsky,
    {Int.\ J.\ Mod.\ Phys.\ D}
   {{\bf 9}, 373 (2000)}
    [{astro-ph/9904398}].
    
    \bibitem{Frieman:2008sn}
    J.~Frieman, M.~Turner and D.~Huterer,
   {Ann.\ Rev.}\
    {Astron.\ Astrophys.\  {\bf 46}, 385 (2008)}
    [{arXiv:0803.0982} [astro-ph]].
    
    \bibitem{Padmanabhan:2002ji}
    T.~Padmanabhan,
    {Phys.\ Rept.\  {\bf 380}, 235 (2003)}
    [{hep-th/0212290}].
    
    \bibitem{Padmanabhan:2006ag}
    T.~Padmanabhan,
    {AIP Conf.\ Proc.\  {\bf 861}, 179 (2006)}
    [{astro-ph/0603114}].
    
    \bibitem{Sahni:2006pa}
    V.~Sahni and A.~Starobinsky,
   {Int.\ J.\ Mod.\ Phys.\ D {\bf 15},}
   {2105 (2006)}
    [{astro-ph/0610026}].
    
    \bibitem{Peebles:2002gy}
    P.~J.~E.~Peebles and B.~Ratra,
    {Rev.\ Mod.\ Phys.\  {\bf 75},}
    {559 (2003)}
    [{astro-ph/0207347}].
    
    \bibitem{Perivolaropoulos:2006ce}
    L.~Perivolaropoulos,
   {AIP Conf.\ Proc.\  {\bf 848}, 698 (2006)}
    [{astro-ph/0601014}].
    
    \bibitem{Sami:2009dk}
    M.~Sami,
    {arXiv:0901.0756} [hep-th].
    
    
    \bibitem{Sami:2009jx}
    M.~Sami,
    Curr.\ Sci.\  {\bf 97}, 887 (2009)
    [{arXiv:0904.3445} [hep-th]].
    
    \bibitem{Sami:2013ssa}
    M.~Sami and R.~Myrzakulov,
    {arXiv:1309.4188} [hep-th].
    
    
    
    \bibitem{Caldwell:1999ew}
    R.~R.~Caldwell,
    Phys.\ Lett.\ B {\bf 545}, 23 (2002)
    [astro-ph/9908168].
    
    \bibitem{Caldwell:2003vq}
    R.~R.~Caldwell, M.~Kamionkowski and N.~N.~Weinberg,
    Phys.\ Rev.\ Lett.\  {\bf 91}, 071301 (2003)
    [astro-ph/0302506].
    
    \bibitem{Carroll:2003st}
    S.~M.~Carroll, M.~Hoffman and M.~Trodden,
    Phys.\ Rev.\ D {\bf 68}, 023509 (2003)
    [astro-ph/0301273].
    
    \bibitem{Singh:2003vx}
    P.~Singh, M.~Sami and N.~Dadhich,
    Phys.\ Rev.\ D {\bf 68}, 023522 (2003)
    [hep-th/0305110].
    
    \bibitem{Hao}
    J.~g.~Hao and X.~z.~Li,
    Phys.\ Rev.\ D {\bf 68}, 043501 (2003)
    [arXiv:hep-th/0305207].
    
    
    
    
    
    
    
    
    \bibitem{Sen:2002nu}
    A.~Sen,
    JHEP {\bf 0204}, 048 (2002)
    [hep-th/0203211].
    
    \bibitem{Sen:2002in}
    A.~Sen,
    JHEP {\bf 0207}, 065 (2002)
    [hep-th/0203265].
    
    \bibitem{Sen:2003mv}
    A.~Sen,
    Phys.\ Scripta T {\bf 117}, 70 (2005)
    [hep-th/0312153].
    
    
    \bibitem{Mazumdar:2001mm}
    A.~Mazumdar, S.~Panda and A.~Perez-Lorenzana,
    Nucl.\ Phys.\ B {\bf 614}, 101 (2001)
    [hep-ph/0107058].
    
    \bibitem{Damour}
  T. Damour, G. W. Gibbons, and C. Gundlach, Phys. Rev. Lett,
      Phys.\ Rev.\ Lett.\  {\bf 64}, 123 (1990)
    [hep-ph/0107058].
    
       \bibitem{Hagiwara}
      Hagiwara K., et al,
    Phys.\ Rev.\ D {\bf 66}, 010001 (2002).
    
    \bibitem{Casas}
    J.A. Casas, J. Garcia-Bellido, and M. Quiros,
    Class.\ Quant.\ Grav.\  {\bf 9}, 1371 (1992).
    
    
    
    \bibitem{Amendola}   
    L.~Amendola,
    Phys.\ Rev.\ D {\bf 62}, 043511 (2000).
    
    	\bibitem{Wetterich} 
      Wetterich C.,
   Astron.\ Astrophys.\  {\bf 301,321},  (1995).
   
       	\bibitem{Farrar} 
  Farrar G. R., Peebles P. J. E.,
    ApJ.\ {\bf 604,1},  (2004).
    
      \bibitem{Gubser}   
   Gubser S. S., Peebles P. J. E.,
    Phys.\ Rev.\ D {\bf 70,12}, 123511 (2004).
    
     \bibitem{Farrar2}
       Farrar G. R., Rosen R. A.,
    Phys.\ Rev.\ Lett.\  {\bf 98,17},  171302 (2007)
   
    
   
      \bibitem{Nonlinear1}   
     A.P. Billyard and A.A. Coley,
    Phys.\ Rev.\ D {\bf 61}, 083503 (2000).
    
     \bibitem{Nonlinear2}   
   N. Bartolo and M. Pietroni, 
    Phys.\ Rev.\ D {\bf 61}, 023518 (2000).
    
    \bibitem{Amendola2004} L.~Amendola, Phys.~Rev.~D, {\bf 69},103524 (2004);
    
    \bibitem{Amendola2008}C.~D.~Porto,~L.~Amendola, Phys.~Rev.~D, {\bf 77}, 083508 (2008);
    
    \bibitem{Piloyan}A.~Piloyan, V.~Marra, M.~Baldi, L.~Amendola, arXiv:1305.3106 [astro-ph].
    
    
    
       
    
    
    
    
       
        \bibitem{L1}   
    aldera-Cabral G.,  Maartens R.,    Urena-Lopez L.~A., 
    Phys.\ Rev.\ D {\bf 79}, 063518 (2009).
     
       \bibitem{L2}   
      Pettorino V.,  Baccigalupi C., 
    Phys.\ Rev.\ D {\bf 77},103003(2008).
    
     \bibitem{L3}   
     Amendola L.,  Baldi M.,    Wetterich C., 
    Phys.\ Rev.\ D {\bf 78},023015(2008).
    
    \bibitem{L4}  Koyama K.,  Maartens R.,    Song Y.-S., JCAP, {\bf 0910}, 017 (2009).
    
     \bibitem{L5}  Caldera-Cabral G.,  Maartens R.,    Schaefer B.~M., JCAP, {\bf 0907}, 027 (2009).
   
     \bibitem{L6}  Valiviita J.,  Majerotto E.,    Maartens R., JCAP, {\bf 0807}, 020 (2008).
     
     
     \bibitem{L7}   
    Didam G. A. Duniya,Daniele Bertacca, Roy Maartens, 
     Phys.\ Rev.\ D {\bf 91},063530(2015).
     
  
   
    
    
    
    
    
   
   \bibitem{Boehmer1}   
   Authors: Christian G. Boehmer, Nicola Tamanini, Matthew Wright, 
  Phys.\ Rev.\ D {\bf 91}, 123002 (2015)).
  
   
  \bibitem{Boehmer2}   
   Authors: Christian G. Boehmer, Nicola Tamanini, Matthew Wright, 
   Phys.\ Rev.\ D {\bf 91}, 123003 (2015)).
  
    
    
         \bibitem{NL1}
     Macci\`{o} A.~V.,  Quercellini C.,  Mainini R.,  Amendola L.,Bonometto
     S.~A., 
    Phys.\ Rev.\ D {\bf 69},123516(2004).
   
    \bibitem{NL2} Baldi M., Mon.~Not.~Roy.~Astron.~Soc., {\bf 411}, 1077 (2011).
   
     \bibitem{NL3} Baldi M.,  {Pettorino} V.,  {Robbers} G.,    {Springel} V., Mon.~Not.~Roy.~Astron.~Soc., {\bf 403}, 1684 (2010).
  
     \bibitem{NL4}
       Li B.,  Barrow J.~D.,
      Phys.\ Rev.\ D {\bf 83}, 024007(2011a).
  
   \bibitem{NL5} Li B.,  Barrow J.~D., Mon.~Not.~Roy.~Astron.~Soc., {\bf 413}, 262 (2011b).
   
   
   \bibitem{NL6} Emma Beynon, Marco Baldi,, David J. Bacon, Kazuya Koyama and Cristiano Sabiu, Mon.~Not.~Roy.~Astron.~Soc., {\bf 422}, 3546–3553 (2012).
   
  
    
    
    
    
    
    
    
    
      \bibitem{C1} Valiviita J.,  Maartens R.,    Majerotto E., Mon.~Not.~Roy.~Astron.~Soc., {\bf 402}, 2355 (2010).
    
     \bibitem{C2}
    Timothy Clemson, Kazuya Koyama, Gong-Bo Zhao, Roy Maartens, Jussi Valiviita
    Phys.\ Rev.\ D {\bf 85}, 043007(2012).
    
    \bibitem{C3}
    Bean R.,  Flanagan E.~E.,  Laszlo I.,    Trodden M.,
    Phys.\ Rev.\ D {\bf 78}, 123514(2008).
    
       \bibitem{C4}  La~Vacca G.,  Kristiansen J.~R.,  Colombo L. P.~L.,  Mainini R.,    Bonometto
       S.~A., JCAP, {\bf 0904}, 007 (2009)
       
       \bibitem{C5}
       Xia J.-Q.,
       Phys.\ Rev.\ D {\bf 80}, 103514(2009).


    \bibitem[\protect\citeauthoryear{Baldi \& Pettorino}{Baldi \&
    	Pettorino}{2010}]{Baldi_Pettorino_2010}
    Baldi M.,  Pettorino V.,  2010, arXiv:1006.3761
    
    
    
    
     \bibitem{LG1}
     Kesden M.,  Kamionkowski M., 
    Phys.\ Rev.\ D {\bf 74}, 083007(2006).
    
       \bibitem{LG2}
     Keselman J.~A.,  Nusser A.,    Peebles P. J.~E., 
    Phys.\ Rev.\ D {\bf 80}, 063517(2009).
      
    \
    
    
    
        \bibitem{IGM1} M. Baldi and M. Viel,  Mon.~Not.~Roy.~Astron.~Soc., {\bf 409}, L89-L93 (2010).
        
        
        
         \bibitem{Bikash}
        Bkash R. Dinda, Anjan A. Sen, 
        Phys.\ Rev.\ D {\bf 97}, 083506(2018).
        
         B. Li and H. Zhao (2010), arXiv:1001.3152 [astro-
ph.CO]].


  \bibitem{Duniya2013}  Didam G.A. Duniya et al
S.~A., JCAP, {\bf 015}, 10 (2013)

    \bibitem{Duniya2016}
Duniya D.
{arXiv:1606.00712} [hep-th].




\bibitem{Kaiser1987} Kaiser N.,  Mon.~Not.~Roy.~Astron.~Soc., {\bf 227}, 1 (1987).


\bibitem{Eisenstein1998} 
D.J.Eisenstein and W.Hu,
ApJ.\ {\bf 496}, 605 (1998).


\bibitem{Sumit}
Sumit Kumar et al ,
Class.\ Quant.\ Grav.\  {\bf 30}, 155011 (2013).




\bibitem{Challinor2011bk}
A.~Challinor and A.~Lewis,
Phys.\ Rev.\ D{\bf 84}, 043516 (2011) [arXiv:1105.5292].

\bibitem{Bruni2011ta}
M.~Bruni, R.~Crittenden, K.~Koyama, R.~Maartens, C.~Pitrou and D.~Wands,
Phys. Rev. D{\bf 85},  041301 (2012) [arXiv:1106.3999].

\bibitem{Jeong2011as}
D.~Jeong, F.~Schmidt and C.~M.~Hirata,
Phys.\ Rev.\ D{\bf 85}, 023504 (2012) [arXiv:1107.5427].

\bibitem{Yoo2010ni}
J.~Yoo,
Phys.\ Rev.\  D{\bf 82},  083508 (2010) [arXiv:1009.3021].

\bibitem{Bartolo2010ec} 
N.~Bartolo, S.~Matarrese and A.~Riotto, 
JCAP {\bf 1104}, 011 (2011) [arXiv:1011.4374]. 

\bibitem{Jeong2011as}
D.~Jeong, F.~Schmidt and C.~M.~Hirata,
Phys.\ Rev.\ D{\bf 85}, 023504 (2012) [arXiv:1107.5427].

\bibitem{Bertacca2012tp} 
D.~Bertacca, R.~Maartens, A.~Raccanelli and C.~Clarkson,
JCAP {\bf 1210}, 025 (2012) [arXiv:1205.5221].

\bibitem{Hu2001yq} 
W.~Hu and A.~Cooray,
Phys.\ Rev.\ D {\bf 63}, 023504 (2001) [astro-ph/0008001].



    
  
\end{thebibliography}
\end{document}